\documentclass{aa}  
\usepackage[varg]{txfonts}

\usepackage{epstopdf}
\usepackage{graphicx}
\usepackage{txfonts}
\usepackage{booktabs}
\usepackage{amsmath}
\usepackage{url}
\bibpunct{(}{)}{;}{a}{}{,}

\begin{document}

   \title{Mid-infrared interferometric variability of \object{DG Tau}: implications for the inner-disk structure\thanks{Based on observations made with the ESO Very Large Telescope Interferometer at Paranal Observatory (Chile) under the 
programs 088.C-1007 (PI: L. Mosoni), 090.C-0040 (PI: Th. Ratzka), and 092.C-0086 (PI: Th. Ratzka).}}
\titlerunning{Mid-infrared variability of DG Tau with VLTI/MIDI}

   \author{J. Varga\inst{1} \and
          K. \'E. Gab\'anyi\inst{1} \and
          P. \'Abrah\'am\inst{1} \and
          L. Chen\inst{1} \and
          \'A. K\'osp\'al\inst{1,4} \and  
          J. Menu\inst{2} \and
          Th. Ratzka\inst{3} \and
          R. van Boekel\inst{4} \and
          C. P. Dullemond\inst{5} \and
          Th. Henning\inst{4} \and
          W. Jaffe\inst{6} \and
          A. Juh\'asz\inst{7} \and
          A. Mo\'or\inst{1} \and
          L. Mosoni\inst{1,8} \and
          N. Sipos\inst{1}
          }
\authorrunning{J. Varga et al.}

   \institute{Konkoly Observatory, Research Centre for Astronomy and Earth Sciences, Hungarian Academy of Sciences, P.O. Box 67, H-1525 Budapest, Hungary \\
              \email{varga.jozsef@csfk.mta.hu}
         \and
                        Instituut voor Sterrenkunde, KU Leuven, Celestijnenlaan 200D, 3001, Leuven, Belgium
\and
Institute for Physics/IGAM, NAWI Graz, University of Graz, Universitätsplatz 5/II, 8010, Graz, Austria
\and
            Max Planck Institute for Astronomy, K\"{o}nigstuhl 17, D-69117 Heidelberg, Germany               
            \and
            Institute for Theoretical Astrophysics, Heidelberg University, Albert-Ueberle-Strasse 2, D-69120 Heidelberg, Germany
            \and
Leiden Observatory, Leiden University, Niels Bohrweg 2, 2333 CA Leiden, The Netherlands
\and
Institute of Astronomy, Madingley Road, Cambridge CB3 OHA, UK
            \and
Park of Stars in Zselic, 064/2 hrsz., H-7477 Zselickisfalud,  Hungary        
             }

   \date{Received; accepted}

 
  \abstract
   {DG Tau is a low-mass pre-main sequence star, whose strongly accreting protoplanetary disk exhibits a so-far enigmatic behavior: its mid-infrared thermal emission is strongly time-variable, even turning the $10~\mu$m silicate feature from emission to absorption temporarily.} 
   {We look for the reason for the spectral variability at high spatial resolution and at multiple epochs.}
   {Infrared interferometry can spatially resolve the thermal emission of the circumstellar disk, also giving information about dust processing. We study the temporal variability of the mid-infrared interferometric signal, observed with the VLTI/MIDI instrument at six epochs between 2011 and 2014. We fit a geometric disk model to the observed interferometric signal to obtain spatial information about the disk. We also model the mid-infrared spectra by template fitting to characterize the profile and time dependence of the silicate emission. We use physically motivated radiative transfer modeling to interpret the mid-infrared interferometric spectra. }
   {The inner disk ($r<1-3$\,au) spectra exhibit a $10~\mu$m absorption feature related to amorphous silicate grains. The outer disk ($r>1-3$\,au) spectra show a crystalline silicate feature in emission, similar to the spectra of comet Hale-Bopp. The striking difference between the inner and outer disk spectral feature is highly unusual among T Tauri stars. The mid-infrared variability is dominated by the outer disk. The strength of the silicate feature changed by more than a factor of two. Between 2011 and 2014 the half-light radius of the mid-infrared-emitting region decreased from
$1.15$ to $0.7$ au.}
   {For the origin of the absorption we discuss four possible explanations: a cold obscuring envelope, an accretion heated inner disk, a temperature inversion on the disk surface and a misaligned inner geometry. The silicate emission in the outer disk can be explained by dusty material high above the disk plane, whose mass can change with time, possibly due to turbulence in the disk. 
}

    \keywords{protoplanetary disks -- stars: pre-main sequence -- stars: individual: DG Tau --  techniques: interferometric -- infrared: stars}

                   \maketitle
%
%
%

\section{Introduction}
\label{sec:intro}

The structure of the inner part of circumstellar disks around pre-main sequence stars determines the initial conditions for terrestrial planet formation. Long-baseline infrared interferometry offers the possibility to achieve the angular resolution required to resolve the innermost regions of planet-forming disks of young stellar objects (YSOs).
The Mid-infrared Interferometric Instrument (MIDI, \citealp{midi}) at the Very Large Telescope Interferometer (VLTI) provided a wealth of information about the size and structure of protoplanetary disks \citep[e.g.,][]{midi_reduc_Leinert,Menu},
and on the spatial distribution of the dust species therein \citep[e.g.,][]{Boekel_silicate}. 
Multi-epoch interferometric sequences could be used to explore changes in the disk structure, enabling study of the temporal physical processes and the dynamics of the inner disk, but this technique has been little explored so far. 
  
\begin{figure}
  \label{fig:DGTau_img}
\resizebox{\hsize}{!}{\includegraphics{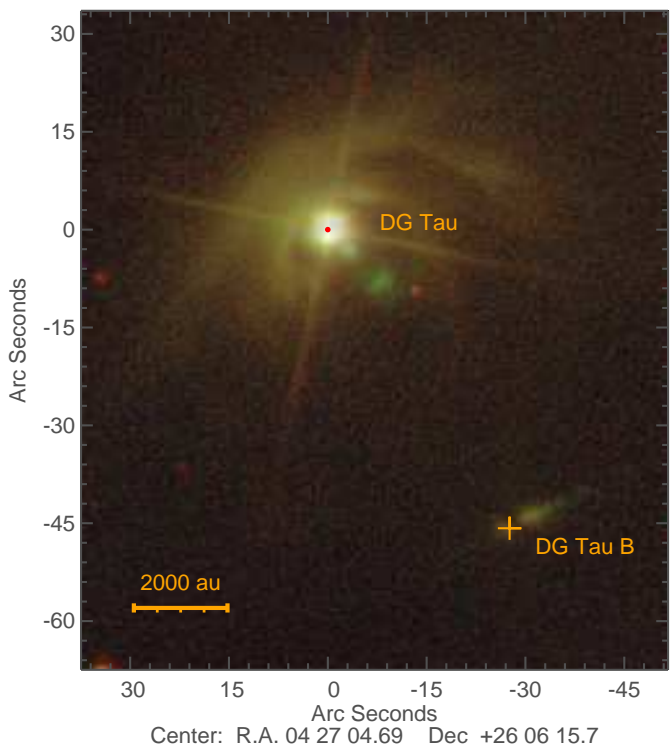}}
                \caption{DG Tau (at 0,0) as seen by the Sloan Digital Sky Survey \citep[SDSS,][]{York2000} in optical wavelengths. The red dot in the center is $1''$ wide, which is the effective field of view of the VLTI/MIDI. An accompanying object, DG Tau B is located at the lower right corner, marked by a cross. This image is a false-color composite created by us using $u$, $g$, $r$, $i$, $z$ images.} \label{fig:DGtau_img}
        \end{figure}

DG\,Tau (see Fig.~\ref{fig:DGTau_img}) is a low-mass K6V pre-main sequence star in the Taurus star forming region, at a distance of about $140$~pc \citep{Ungerechts1987}, with a stellar luminosity of $0.9$~L$_\sun$ \citep{Palla2002}. It is a classical T Tauri star with a significant accretion rate \citep[$\dot M = 4.6 \times 10^{-8} - 7.4 \times 10^{-7}$ M$_\sun$ yr$^{-1}$,][]{WhiteGhez2001,WhiteHillenbrand2004}. It also has a well studied jet with knots and bow shocks extending to at least $1500$~au \citep[][and references therein]{Gudel2008,Maurri2014}. \cite{inclination} performed observations of DG Tau with the Combined Array for Research in Millimeter-wave Astronomy (CARMA) at $1.3$\,mm and at $2.7$\,mm. Modeling these observations, they found an outer radius of $70-80$\,au, an inclination angle of $i=28^\circ \pm 10^\circ$, a disk position angle of $\phi=135^\circ \pm 21^\circ$, and a disk mass between $0.009$ and $0.07$ $\mathrm{\,M}_\sun$ depending on the model parameters. 


DG Tau is known to have dramatic variability in its $10~\mu$m silicate feature \citep[e.g.,][and references therein]{Bary_spitzer}. A unique aspect of this variability is that the silicate feature was detected in absorption as well as emission, and during a few epochs it could not be detected at all. Although  several models have been proposed \citep[e.g.,][]{Bary_spitzer, sitko1}, the origin of variability is not yet understood.

Here we present a study of the variability of the silicate feature in DG Tau, based on our multi-epoch interferometric data set. Our aim is to investigate the spatial distribution of dust species in the disk, the variation of the silicate feature at different spatial scales, and the temporal variability of the inner disk. In Sect.~\ref{sec:obs} we describe the observations and the data reduction. In Sect.~\ref{sec:res} we show the results of the interferometric modeling, estimate disk size for each epoch, and model mid-infrared spectra by template fitting. In Sect.~\ref{sec:discuss} we delineate and discuss several physical processes which can explain our findings. Finally, in Sect.~\ref{sec:concl}, we summarize our results. 
  

\section{Observations and data reduction}
\label{sec:obs}

\subsection{Observations}

\begin{table*}
\caption{Overview of VLTI/MIDI observations of DG Tau. $B$ is the projected baseline length and $\phi_B$ is the projected position angle of the baseline (measured from North through East). The resolution is the approximate diameter of the beam at $10.7~\mu$m, converted to physical scale. In the last two columns we list the name of the calibrators and the time of their measurement. 
}             
\label{table:obs}      
\begin{center}                          
\begin{tabular}{l l l l l l l l l}        
\toprule \toprule               
\multicolumn{7}{c}{Target} & \multicolumn{2}{c}{Calibrator} \\    
\cmidrule(lr){1-7} \cmidrule(lr){8-9}
Date and time & Telescopes & $B$  &$\phi_B$ & Resolution  & Seeing & Airmass & Name & Time  \\    
(UTC) & & (m) & ($^\circ$) & (au) & (\arcsec) & & & (UTC) \\
\midrule                   

2011-10-10 06:37 & U1-U2 & 33 & 30 & 5.7 & 0.8& 1.7 & HD 27482 & 06:51\\
2011-10-10 07:43 & U1-U3 & 76 & 45 & 2.5 & 0.8& 1.6 & HD 27482 &  07:29\\2011-12-13 01:53 & U1-U3 & 56 & 36 & 3.4 & 0.8\tablefootmark{a}& 1.8& HD 27482 & 01:32\\
2011-12-13 02:59 & U1-U2 & 37 & 34  & 5.1 & 0.7& 1.6& HD 27482 & 02:38\\
& & & & & & & HD 27482 & 03:19\\
2012-02-04 01:02\tablefootmark{b} & U1-U2 & 45 & 40 & 4.2 & 1.5& 1.6& HD 27482 & 00:42\\
& & & & & & & HD 27482 & 01:23\\
2012-02-05 00:46 & U1-U2 & 44 & 39 & 4.3 & 1.1& 1.6& HD 27482 & 00:29\\
& & & & & & & HD 27482 & 01:02\\
2012-02-06 00:39\tablefootmark{c} & U1-U3 & 83 & 46 & 2.3 & 0.9& 1.6& HD 27482 &  00:22\\
2012-11-03 05:07 & U1-U3 & 64 & 41 & 2.9 & 0.9 & 1.7& HD 27482 & 04:29\\
2014-01-14 02:39 & U1-U3 & 88 & 45 & 2.1 & 0.8& 1.7& HD 20644 & 01:17\\
& & & & & & & HD 25604 & 02:27\\
2014-01-15 01:47 & U2-U4 & 89 & 83 & 2.1 & 1.5& 1.6 & HD 20644 & 00:42\\
& & & & & & & HD 25604 & 02:43\\
& & & & & & & HD 20644 & 02:03\\
2014-01-15 02:17 & U2-U4 & 89 & 80 & 2.1 & 0.9& 1.6& HD 20644 & 00:42\\
& & & & & & & HD 25604 & 02:43\\
& & & & & & & HD 20644 & 02:03\\
2014-01-15 02:22 & U2-U4 & 89 & 80 & 2.1 & 0.7& 1.6& HD 20644 & 00:42\\
& & & & & & & HD 25604 & 02:43\\
& & & & & & & HD 20644 & 02:03\\

\bottomrule                               
\end{tabular}
\end{center}
\tablefoot{
\tablefoottext{a}{The seeing for this epoch was missing from the observation data. We indicate here an interpolated value.}
\tablefoottext{b}{According to the ESO observing log, this observation had to be repeated the following day because of its poor quality. Therefore we do not use these data in our work. }
\tablefoottext{c}{The total flux observation was not of adequate quality so it was discarded.}
}
\end{table*}

The VLTI/MIDI instrument \citep{midi} combines the signal from two 8.2 m Unit Telescopes (UT) or from two 1.8 m Auxiliary Telescopes (AT). It provides spectrally resolved interferometric data in the wavelength range of $7.5-13~\mu$m. Between 2011 and 2014, we performed a dedicated program by conducting interferometric observations of DG Tau with the VLTI/MIDI using UTs in several epochs. 
All observations were performed in HIGH-SENS mode with the low-resolution ($\lambda /  \Delta\lambda \approx30$) prism \citep{midi_reduc1,midi_reduc_Ratzka}. The measurements were taken at various projected baseline lengths ($B$) and position angles ($\phi_B$), as summarized in Table \ref{table:obs}. The values of the position angles fall into two narrow intervals, one between $30^\circ$ and $45^\circ$ and one $\sim$\,80$^\circ$. The projected baseline lengths range from $33$\,m to $89$\,m. Fig. \ref{fig:uv_sizes} (panel a) shows the distribution of the baselines in the $uv$-plane. Since most observations were taken with similar position angles but with different baseline lengths (U1-U2 and U1-U3 configurations), we are able to sample the disk at $30^\circ$ to $45^\circ$ on different spatial scales. Additionally, we have high-resolution spatial information at a direction of $\sim$\,80$^\circ$ (U2-U4 configuration).

In all epochs, the obtained data sets consist of the N-band ($7.5-13~\mu$m) low-resolution spectrum (hereafter total spectrum, $F_\mathrm{tot,\nu}$), and the interferometric spectrum (hereafter correlated spectrum, $F_\mathrm{corr,\nu}$) of the target, both in the same wavelength range with the same spectral resolution. Visibility is defined as $V =  F_\mathrm{corr,\nu} / F_\mathrm{tot,\nu}$. There is one exception, on 2014 January 15: Correlated spectra were taken in three consecutive time slots, but only two total spectra were recorded. The majority of our observations have good quality. The observations taken on 2012 February 4 had to be repeated the following night; according to the observation log the errors were too high. Therefore we discard this observation from the further analysis. On 2012 February 6, the total spectrum was not of adequate quality, thus it is also excluded.

\subsection{Data reduction} 

For the interferometric data reduction we used the Expert Work Station (EWS) package 2.0\footnote{The EWS package can be obtained from: \url{home.strw.leidenuniv.nl/~nevec/MIDI/index.html}}, which is based on a coherent linear averaging method \citep{midi_reduc1,faintMIDI}, and is one of the standard tools for processing MIDI data. 
EWS routines were called from a python wrapper developed by \cite{Menu}. 

We applied the ``direct flux calibration'' method as described in \cite{faintMIDI} instead of visibility calibration. In this way one can avoid the usage of the less accurate total spectrum measurements when calibrating the correlated spectrum of the target. The higher uncertainty level of the total spectrum is due to the strong and variable background in the MIDI wavelength range. When observing the interferometric signal, the incoherent background noise cancels out. However, to subtract the background in the total spectrum chopping was used, which may leave residual sky signal. 

The calibrated correlated spectrum of the target can be obtained from the observed (raw) correlated spectrum of the target by dividing it with a transfer function, which following \cite{Menu}, can be written as:
\begin{equation}
T_\mathrm{corr}=\frac{C^\mathrm{cal}_\mathrm{raw}}{F^\mathrm{cal} V^\mathrm{cal}},
\label{transfer_func}
\end{equation}
where $C^\mathrm{cal}_\mathrm{raw}$ is the observed (raw) correlated spectrum of the calibrator source, $F^\mathrm{cal}$ is the known total spectrum of the calibrator, and $V^\mathrm{cal}$ is the visibility of the calibrator (calculated from its known diameter). $V^\mathrm{cal}$ is very close to unity, since typically point-like, unresolved sources are chosen as MIDI calibrators. To determine $T_\mathrm{corr}$, it is preferable to use all calibrators observed on the same baseline as our target in a given night. Only in the last epochs could we use more than one calibrator. In the previous epochs we used $1-2$ measurements of the same calibrator (Table \ref{table:obs}).
$T_\mathrm{corr}$ is a time- and airmass-dependent quantity. Using the routines of \citet{Menu} the time-dependence of $T_\mathrm{corr}$ was taken into account by applying linear interpolation. To correct for airmass ($X$) dependency, $T_\mathrm{corr}$ values were multiplied by the correction factor $\exp(f_\lambda X)$, where $f_\lambda$ is the wavelength dependent atmospheric extinction. Calibration of the total spectrum of our target was conducted in a similar way as the correlated spectrum, but using $V^\mathrm{cal} \equiv 1$ in Eq. (\ref{transfer_func}). In this case the transfer function is the atmospheric transparency.

The outputs of the data reduction pipeline are the calibrated total spectrum and the calibrated correlated spectrum of the target. Visibilities are also calculated as the correlated spectrum divided by the total spectrum. The correlated spectrum can be interpreted as the spectrum of the very inner part of an object. In the case of DG Tau this means that we have spectral information with a resolution 
of $2 - 6$~au, depending on the baseline length (see Table \ref{table:obs}). 

The accuracy of the total spectra is not adequate to detect variations from one night to the next, therefore we averaged the calibrated total spectra taken on consecutive nights to increase the signal-to-noise ratio. This resulted in five total spectra: 2011 October, 2011 December, and 2012 February, obtained by averaging two measurements; 2014 January, by averaging three measurements; and 2012 November, when only one total spectrum was measured. Because of the atmospheric ozone absorption feature, data points between $9.4~\mu$m and $10.0~\mu$m have significantly higher uncertainties than in other parts of the total spectrum, therefore we decided to discard points of the total spectra in this wavelength regime from the analysis.

In the case of correlated spectra, we average only those observations that have the same baseline configuration.
On 2014 January 15 three consecutive correlated spectra were observed with the same baseline length, and position angles within a few degrees, so we averaged these measurements. Thus we ended up with nine independent correlated spectra (see Table \ref{table:corr_fit}).

\subsection{Calibration quality}
\label{sec:calqual}

MIDI spectra are known to have calibration biases \citep{faintMIDI}. The uncertainties on the spectra provided by the EWS seem to be much larger than the variance of the neighboring spectral data points (see Fig. \ref{fig:uncorr_matrix} for example). This suggests that the main error source is systematic, due to calibration issues, that is, the absolute level of the spectrum can only be calibrated with a higher uncertainty, although the spectral shape is more reliable. The calibration biases arise from flux losses in the MIDI instrument \citep{faintMIDI}. 

To verify the calibration quality, we reduced calibrators of DG Tau as if they were science targets, using other calibrators to calibrate them. A calibrator is supposed to be non-variable, and compact, so it remains mostly unresolved by the interferometer. From Table \ref{table:obs} we can see that all observations from 2011 and 2012 were done using the same source (HD 27482) for calibration. Data from 2014 were calibrated with two other stars (HD 20644, HD 25604). We obtained five correlated spectra, and three total spectra for HD 27482. The standard deviation of these spectra at $10.7~\mu\mathrm{m}$ is $0.4$~Jy ($7 \%$) both for the correlated and total flux density. HD 20644 was observed three times on two consecutive days. The standard deviation for the correlated flux densities (again at $10.7~ \mu\mathrm{m}$) is $1.7$~Jy ($10 \%$), while for the total flux densities one finds $2.0$~Jy ($12 \%$). 

We take these standard deviations as the errors of the absolute calibration. The error bars on the calibrator spectra calculated by the reduction pipeline are relatively similar: $7 - 10 \%$ for the correlated spectra and $10 - 14 \%$ for the total spectra. Thus we conclude that the uncertainties provided by the EWS are reliable. In the case of DG Tau, typical measurement errors are $\sim$\,$ 6 \%$ ($\sim$\,$0.15$~Jy at $10.7~ \mu\mathrm{m}$) for the correlated spectra and $10 - 20 \%$ ($0.5 - 1.0$~Jy at $10.7~ \mu\mathrm{m}$) for the total spectra.






\section{Results}
\label{sec:res}

  \subsection{Interferometric modeling}
\label{sec:int_model}
The visibilities and correlated spectra contain spatial information about our  target. Because of the sparse $uv$-sampling and lack of closure phases, we cannot obtain an image of the disk, instead we can try to fit simple disk models to the interferometric data. Note that we use correlated spectra for the fitting, instead of visibilities, because the latter have much larger measurement uncertainties.

The aim of the model fitting is to measure the size of the mid-infrared emitting region of the DG Tau disk. We fit the measured correlated flux densities with a physically motivated geometric disk model also used by \citet{Menu}. For the disk inclination and position angle we adopted the values of \citet{inclination} (Sect. \ref{sec:intro}).

The model geometry is a thin, flat disk, which starts at the dust sublimation radius and extends out to $R_\mathrm{out} = 300$~au, where the mid-IR radiation is negligible. The temperature drops as a power-law, and the disk emits thermal radiation:
\begin{equation}
I_\nu\left(r\right) = \tau_\nu B_\nu\left(T\left(r\right)\right),
\end{equation} 
where $\tau_\nu$ is the optical depth, $B_\nu$ is the Planck-function and the temperature ($T$) as a function of radius ($r$) is given as
\begin{equation}
T\left(r\right) = T_\mathrm{sub}\left(\frac{r}{R_\mathrm{sub}}\right)^{-q}.
\end{equation}
Here $T_\mathrm{sub}$ is the dust sublimation temperature, fixed at $1500$~K. $R_\mathrm{sub}$ is the sublimation radius, calculated from the known luminosity ($L_\star$) of the central star:
\begin{equation}
R_\mathrm{sub} = \left(\frac{L_\star}{4\pi \sigma T_\mathrm{sub}^4}\right)^{1/2}.
\end{equation}
With $L_\star = 0.9$~L$_\sun$, the sublimation radius is $R_\mathrm{sub} = 0.07~\mathrm{au}$. The model has two free parameters: $q$, which is the power-law exponent of the temperature profile, and $\tau_\nu$, the optical depth. The latter can be related to the total flux density of the object ($F_\mathrm{tot,\nu}$), which is also measured by MIDI, as follows:

\begin{equation}
F_\mathrm{tot,\nu} = \int_{R_\mathrm{sub}}^{R_\mathrm{out}} 2 \pi r \tau_\nu B_\nu\left(T\left(r\right)\right) \mathrm{d}r.
\end{equation} 
$F_\mathrm{tot,\nu}$ only includes the flux from the disk. At mid-infrared wavelengths the stellar photospheric flux is negligible. We estimate that the stellar flux is at most $3 \%$ of the disk flux, also taking into account the possible contribution from accretion hot spots on the stellar surface.

At a specific wavelength, we have correlated flux densities measured as a function of the projected baseline length. The brightness profile of the model disk should be transformed in the $uv$-space. This transformation is known as a Hankel transform, which is essentially the two-dimensional Fourier transform of a circularly symmetric function, with the following formula:
\begin{equation}
F_\mathrm{corr,\nu}\left(B_{\mathrm{eff}}\right) = 
F_\mathrm{tot,\nu} 
\frac
{\int_{R_\mathrm{sub}}^{R_\mathrm{out}} r B_\nu\left(T\left(r\right)\right) J_0\left(2\pi r B_{\mathrm{eff}} / \left( \lambda d \right) \right) \mathrm{d}r }
{\int_{R_\mathrm{sub}}^{R_\mathrm{out}} r B_\nu\left(T\left(r\right)\right)  \mathrm{d}r }.
\end{equation}
Here $J_0$ is the 0th order Bessel function, $d$ is the distance from the source and $B_{\mathrm{eff}}$ is the effective projected baseline length, corrected for the inclination ($i$) and position angle ($\phi$) of the disk as follows:
\begin{equation}
\begin{split}
\alpha = \mathrm{atan2}\left(\sin\left(\phi_B-\phi\right),\cos i \cos\left(\phi_B-\phi\right)\right),
\end{split}
\end{equation}
\begin{equation}
\begin{split}
B_{u, \mathrm{eff}} = B \cos \alpha \cos \phi - B \cos i \sin \alpha \sin \phi,
\end{split}
\end{equation}
\begin{equation}
\begin{split}
B_{v, \mathrm{eff}} = B \cos \alpha \sin \phi + B \cos i \sin \alpha \cos \phi,
\end{split}
\end{equation}
\begin{equation}
B_{\mathrm{eff}} = \sqrt{B_{u, \mathrm{eff}}^2 + B_{v, \mathrm{eff}}^2}.
\end{equation}

\begin{figure}
\resizebox{\hsize}{!}{  \includegraphics{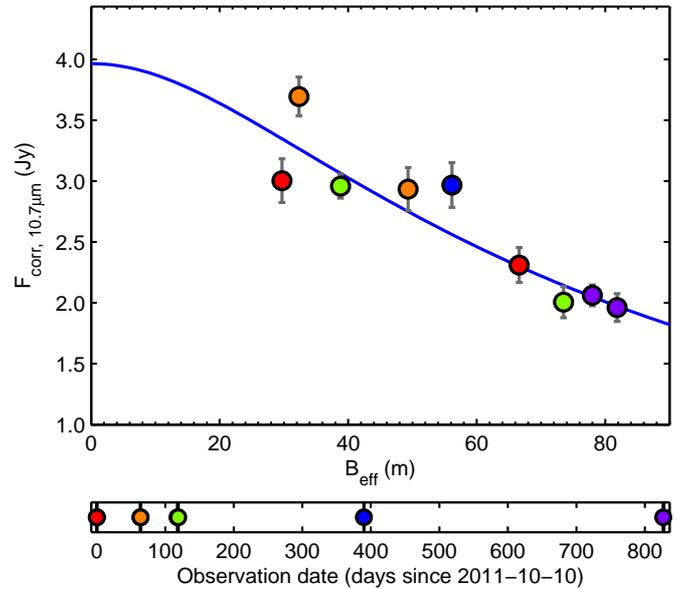}}
          \caption{$10.7~\mu$m correlated flux densities as a function of the effective baseline length ($B_\mathrm{eff}$, see text) for DG Tau, showing the result of the model fitting (blue curve). The symbols are color coded for observation date.} 
        \label{fig:model}
        \end{figure}

Fig.~\ref{fig:model} shows a fit to our data at $\lambda = 10.7~\mu\mathrm{m}$, obtained by minimizing the reduced $\chi^2$ using the measurement uncertainties provided by the data reduction pipeline.
The exponent $q$ determines how rapidly the disk fades outwards, thus it can be used as a measure of the compactness of the emission. The best-fit value for $q$ is $0.54 \pm 0.02$, which closely matches what one would expect for the optically thin emission from a warm disk surface layer. Following \citet{Menu} we define $r_\mathrm{e}$, a half-light radius:
\begin{equation}
\frac{F_\mathrm{tot,\nu}}{2} = 
\int_{R_\mathrm{sub}}^{r_\mathrm{e}} 2 \pi r I_\nu\left(r\right) \mathrm{d}r.
\end{equation}
\begin{table}
\caption{Results of the interferometric modeling, for three wavelengths, calculated from the correlated spectra.
}             
\label{table:hlr}      
\begin{center}                          
\begin{tabular}{l l l l l}        
\toprule \toprule                
$\lambda$  & $F_{\mathrm{tot},\nu}$  & $q$ & $r_\mathrm{e}$ & $r_\mathrm{e}$  \\    
($\mu$m) &  (Jy) & & (mas) & (au)\\
\midrule                      
$~8.0$ & $3.3 \pm 0.1 $ & $0.58 \pm 0.03 $ & $2.5 \pm 0.3 $ & $0.35 \pm 0.05$\\ 
$10.7$ & $4.0 \pm 0.2$ & $0.54 \pm 0.02 $ & $5.2 \pm 0.4$ & $0.73 \pm 0.10$ \\ 
$13.0$ & $5.8 \pm 0.4$ & $0.53 \pm 0.02$ & $8.2 \pm 0.9$ & $1.15 \pm 0.19$\\ 
\bottomrule 
\end{tabular}
\end{center}
\end{table}

    From the best fit in Fig.~\ref{fig:model}, the resulting half-light radius at $10.7~\mu$m is $r_\mathrm{e} = 0.73 \pm 0.10$~au (see Table~\ref{table:hlr} for results for other wavelengths). The disk appears more extended with increasing wavelength (Table~\ref{table:hlr}). This is because the bulk of the radiation at larger wavelengths comes from a wider, cooler area with a larger average distance from the star \citep{Schegerer2008}. 

Due to the high accretion rate, mentioned in Sect.~\ref{sec:intro}, the accretion luminosity of DG Tau can be also significant. Thus the total luminosity of the source can be higher than the luminosity of the stellar photosphere, which can move $R_\mathrm{sub}$ outwards.  The variable accretion rates yield $1-10~\mathrm{L}_\sun$ accretion luminosities, and $R_\mathrm{sub}$ will be in the range of $0.1-0.2$~au. There is a $40\%$ increase in $q$ with $10~\mathrm{L}_\sun$ total luminosity, with respect to the photosphere-only case. With a modest accretion luminosity of $1~\mathrm{L}_\sun$ the change in $q$ is only $10\%$. However, the resulting $r_\mathrm{e}$ values are completely insensitive to these changes in $R_\mathrm{sub}$, changing by only $4\%$ at $10.7~\mu$m.

The scatter of the correlated flux densities along the fitted curve in Fig.~\ref{fig:model} is $ 10-20 \%$. It means that a single non-variable disk model can constrain the average size of the mid-infrared emitting region.
However, the correlated flux densities measured at the two shortest baselines ($B_\mathrm{eff} < 40$~m) show
 $\sim$\,$20 \%$ flux difference over two months, thus hinting at temporal variability. In the following section, we elaborate on this possible size variability.

\subsection{Size variability}
\label{sec:size_var}

    \begin{figure*}
    \centering
        \includegraphics[width=17cm]{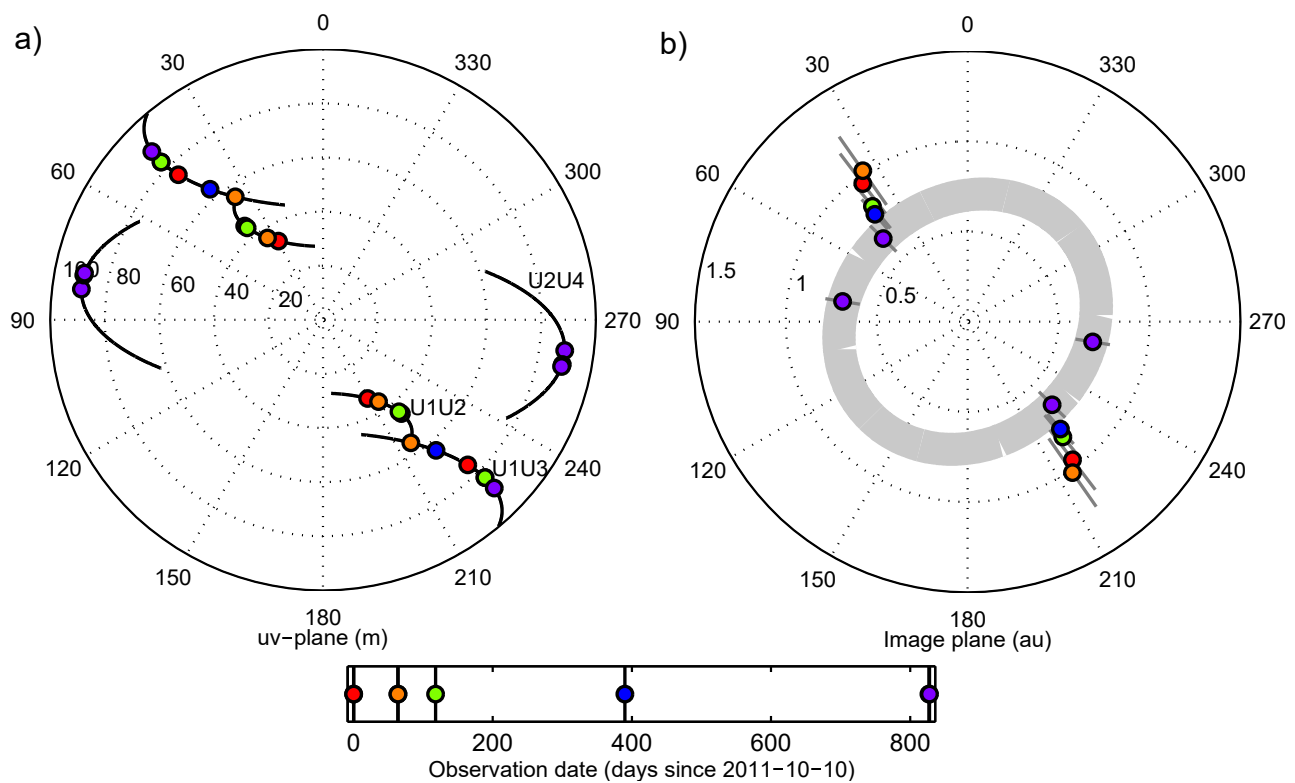}

          \caption{a) $uv$-plot of DG Tau MIDI observations. The tracks correspond to specific telescope configurations. b) Disk half-light radii, determined from visibilities for each epoch. The symbols are color-coded for observation date. Position angles are measured from North to East (counterclockwise). The radial scale of the plot is given in au.}
        \label{fig:uv_sizes}
        \end{figure*}

\begin{table}
\caption{Results of the interferometric modeling calculated from the visibilities for each epoch (for $\lambda = 10.7~\mu\mathrm{m}$). 
}             
\label{tab:sizes}      
\begin{center}                          
\begin{tabular}{l l l l l}       
\toprule \toprule                
Epoch & $\phi_B$ & $q^V$ & $r_\mathrm{e}^V$ & $r_\mathrm{e}^V$ \\ 
    & ($^\circ$) & & (mas) & (au) \\
\midrule                       
2011 Oct & 37 & $0.50 \pm 0.02$ & $7.7 \pm 1.3$ & $1.08 \pm 0.20$ \\
2011 Dec & 35 & $0.49 \pm 0.02$ & $8.2 \pm 1.7$ & $1.15 \pm 0.28$ \\
2012 Feb & 39 & $0.51 \pm 0.03$ & $6.7 \pm 1.3$ & $0.94 \pm 0.22$ \\
2012 Nov & 41 & $0.52 \pm 0.02$ & $6.3 \pm 0.6$ & $0.89 \pm 0.13$ \\
2014 Jan & 45 & $0.54 \pm 0.02$ & $5.2 \pm 0.7$ & $0.73 \pm 0.12$ \\
2014 Jan & 80 & $0.53 \pm 0.02$ & $5.4 \pm 0.6$ & $0.76 \pm 0.13$ \\
\bottomrule 
\end{tabular}
\end{center}
\end{table}

In the previous section, we deduced the average disk size using all the correlated spectra. Nevertheless, we cannot determine individual sizes for each epoch due to the limited range of observed baselines on a given night. However, the total spectrum measurements (corresponding to the source at zero baseline length) can be used to determine sizes for each epoch separately. Such calculations are generally less accurate than correlated flux measurements at multiple baselines, but still enable us to study time variability of the disk size.


We utilize here almost the same model as in Sect.~\ref{sec:int_model}, but we use visibilities measured at $\lambda = 10.7~\mu\mathrm{m}$ instead of correlated flux densities. 
The resulting disk sizes are shown in Table~\ref{tab:sizes} and in Fig.~\ref{fig:uv_sizes} (panel b). It is a polar plot showing the measured sizes at specific baseline position angles. Note that the values in the table are corrected for inclination, but the plotted values in the figure are not.
Assuming that there was no significant change in the brightness distribution from one night to another, we can make some constraints on the disk inclination using the size values from the last two epochs (shown in purple in Fig.~\ref{fig:uv_sizes} panel b). The gray ellipse on the plot shows the shape of the projected disk from mm observations (\citealp{inclination}, arbitrarily scaled). From the figure we can readily see that our mid-IR shape is consistent with the millimetric shape within the uncertainties.


With the sizes determined for each epoch, we can now look for the signs of temporal size variations. Although the values have relatively large error bars, there seems to be a decreasing trend in the disk size: between 2011 and 2014 the half-light radius decreased from $r_\mathrm{e}^V =  1.15\pm 0.28$~au to $0.73\pm 0.12$~au. We performed a linear fit to the sizes as a function of time, and the resulting rate  of change is $-0.14 \pm 0.07$ au$/$yr, which is a $2 \sigma$ result for variability. In parallel, total flux also seems to delineate a decreasing trend. The size at the last epoch is consistent with the average size derived from the correlated spectra in the previous section. This means the whole disk can be described with a single temperature component. At earlier epochs, however, $r_\mathrm{e}^V > r_\mathrm{e}$, which suggest that the outer disk may have a shallower temperature profile. This may imply an ongoing structural rearrangement of the dusty disk material.
\citet{2016Millan-Gabet} observed DG Tau with the Keck Interferometer in 2010, and got $1.15\pm 0.02$~au for the mid-infrared (N-band) size by fitting a Gaussian model with an inner gap. This value fits the trend we observed. We note, however, that their value may not be directly comparable to our sizes due to the different fitting techniques.   
    
From the interferometric modeling in Sect. \ref{sec:int_model} we have seen, that the difference of individual correlated flux densities from the best-fit model is in the range of $10-20 \%$. This implies that real variability of the half light radius derived from the correlated flux densities should be in the same order. However, indicated by the half-light radii from the visibilities (i.e., including the total flux density) we can observe somewhat larger size variations (up to $30 \%$) among the different epochs, indicating  that the total spectrum is much more variable than the correlated spectrum.
In the next section we show a detailed analysis of the mid-infrared spectral behavior of DG Tau.

\subsection{Mid-infrared spectral behavior of DG Tau}

DG Tau is famous for its highly variable silicate feature in its mid-infrared spectrum \citep[e.g.,][]{silicate_woodward,kospal_atlas}. As early as in 1982 and 1983, \cite{85abs} observed the mid-infrared spectrum of DG Tau and found that it showed the silicate feature in absorption. Later,  \cite{silicate_wooden} reported that it showed silicate in emission
in 1996 \citep{silicate_wooden}. At that time, amorphous and crystalline forms of olivine and pyroxene were identified, while in the following year (in 1997 September) the mineralogy was dominated by Mg-rich crystalline olivine. In 1998, the silicate feature turned into absorption, and a drop in visual magnitude was observed prior to this event \citep{silicate_wooden}. 
Although with different shape, the silicate feature was also in absorption in 2001, and this was simultaneous with an overall brightness increase in the $3-5~\mu$m wavelength range. The feature disappeared by 2001 October \citep{silicate_woodward}.
A {\it Spitzer} spectrum in 2004 \citep{kospal_atlas} indicates a weak silicate emission. According to \cite{sitko1}, in 2006, DG Tau again experienced an outburst in the $3-5~\mu$m range and the silicate feature appeared in emission in its spectrum again. \cite{Bary_spitzer} presented nine {\it Spitzer} observations of DG Tau obtained between 2004 and 2007. The silicate feature showed variations over month- and year-long timescales; shorter variations on day- to week-long timescales were not detected. The spectra were dominated by crystalline forsterite in emission in all epochs. Recently, \cite{2016Millan-Gabet} reported mid-infrared and near-infrared measurements of DG Tau performed in 2010 September, October, and November with the Keck Interferometer. At that time, the spectrum did not show a silicate emission feature. 

\begin{figure}
        \resizebox{\hsize}{!}{\includegraphics[angle=-90, clip=]{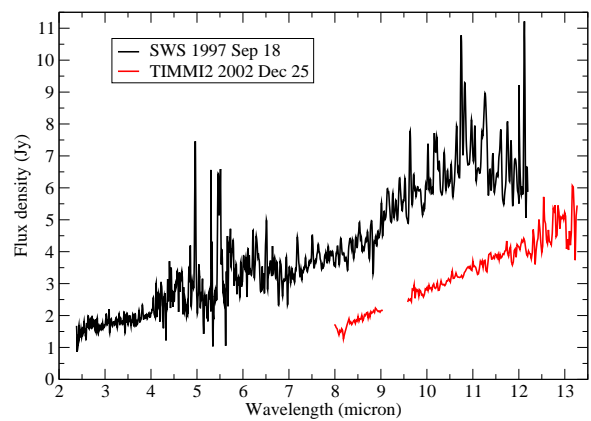}}
        \caption{DG Tau mid-infrared spectra taken by the SWS on 1997 September 18 \citep{ISO_archive_DGTAU} and by TIMMI2 on 2002 December 25 \citep{P_phd}. From the latter the wavelength range affected by the atmospheric ozone feature ($\sim9-10$\,$\mu$m) is excluded. \label{fig:timmi2}}
\end{figure}

Checking the literature, we identified two additional observations of the silicate feature of DG Tau \citep[see Fig. \ref{fig:timmi2},][]{ISO_archive_DGTAU,P_phd}. On 1997 September 18, a spectrum taken by the Short Wavelength Spectrometer (SWS) on-board the Infrared Space Observatory\footnote{Data were downloaded from the ISO archive \url{http://iso.esac.esa.int/ida/}} showed weak emission. This is consistent with the report of \cite{silicate_wooden} on the detection of a similar silicate emission feature in 1997 September. The other observation taken by the TIMMI2 instrument at La Silla Observatory on 2002 December 25 showed no feature, similarly to the 2001 October data. This might be indicative of a longer featureless period.

\begin{figure*}
\centering
        \includegraphics[angle=-90, width=17cm,trim={0cm 1cm 1.5cm 2.5cm}, clip]{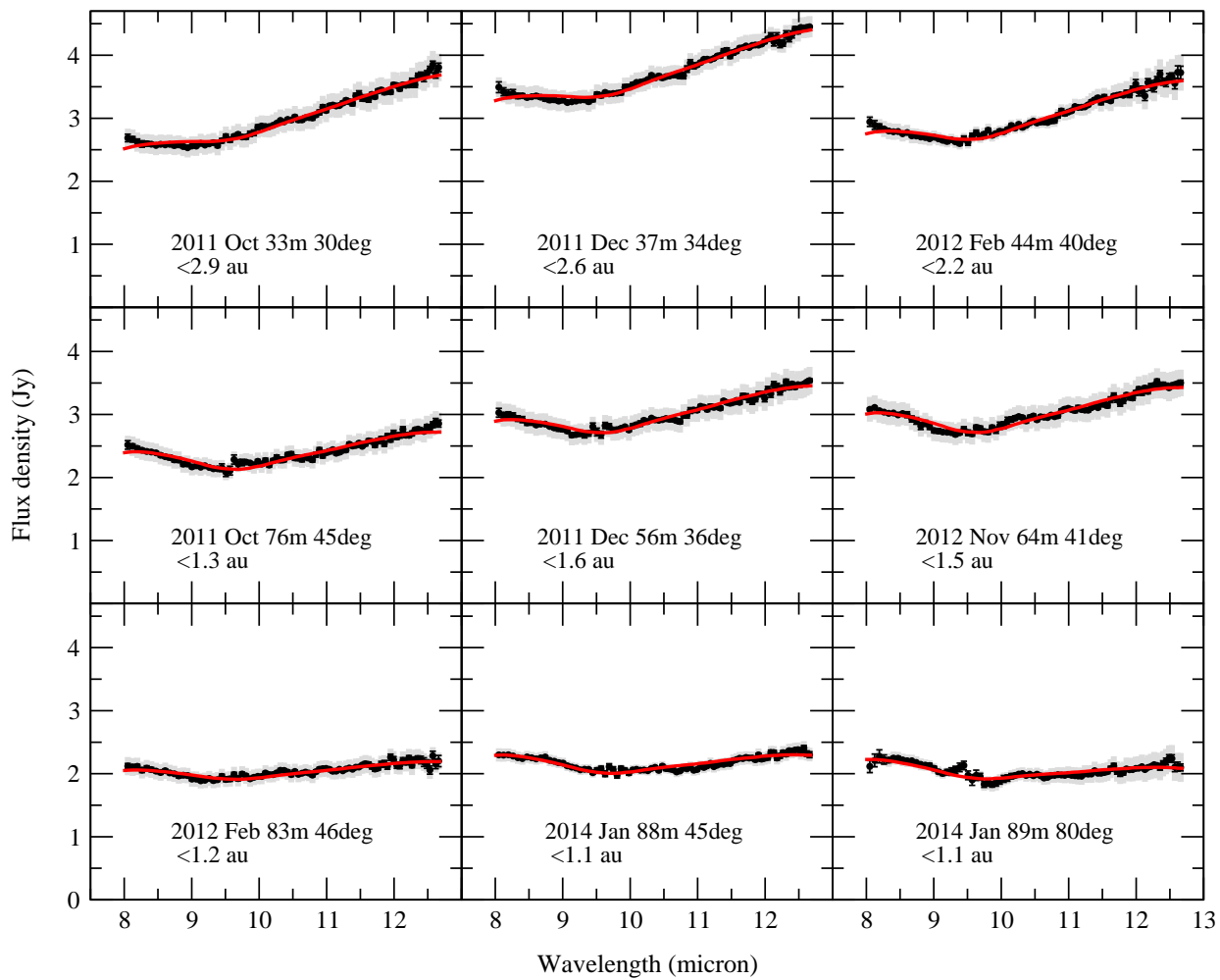}
        \caption{Correlated spectra (black filled circles) and best fit curves (solid red line) of DG Tau at different projected baselines and in different epochs. The rows are categorized by approximate size of the observed baseline (top: short baseline, middle: intermediate baseline, bottom: long baseline). Gray shading indicates the level of total calibration uncertainties, while black error bars represent the random point-to-point uncertainties (see Appendix \ref{app:err} for more details on error calculation).\label{fig:corr_matrix}}
\end{figure*}
       
       \begin{figure}
\resizebox{\hsize}{!}{\includegraphics{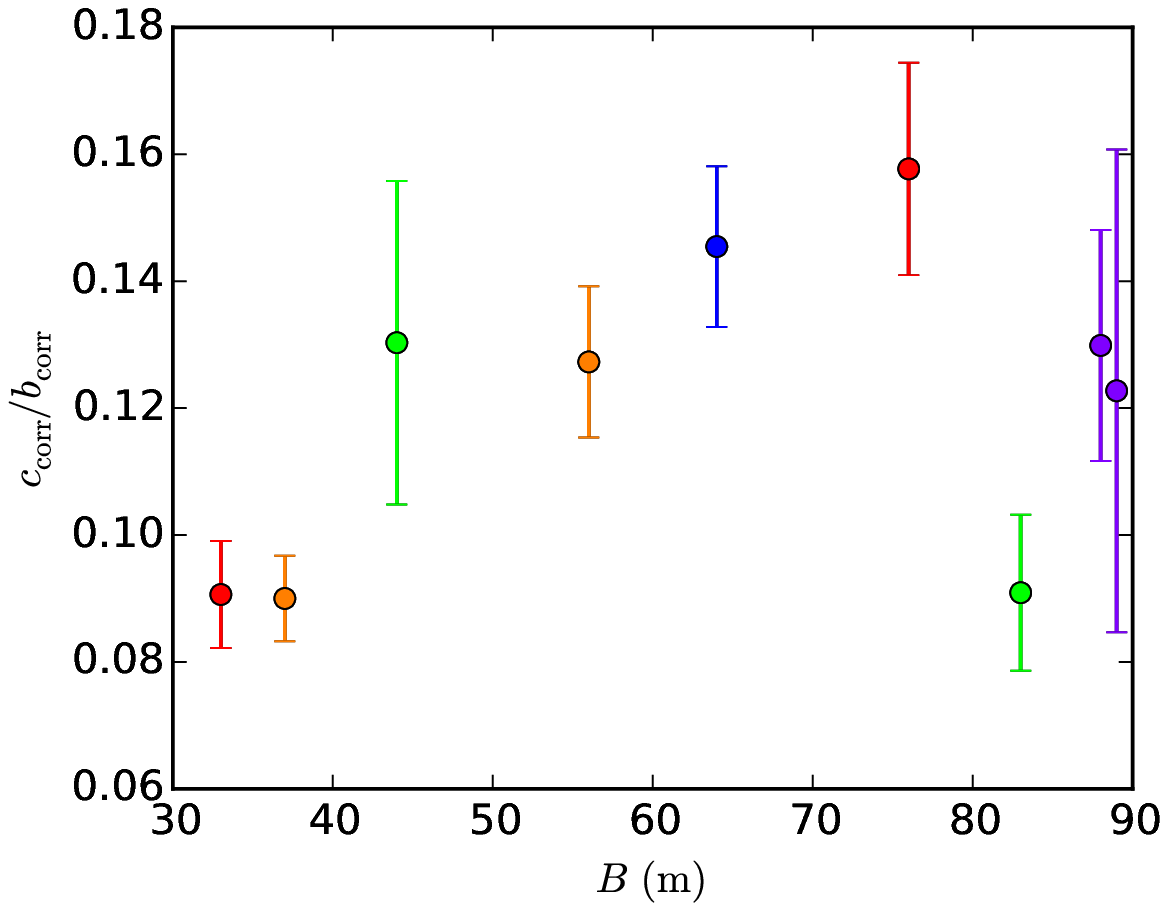}}
                \caption{Amplitude of the silicate absorption feature ($c_\mathrm{corr}$) normalized for the $10.7\,\mu\mathrm{m}$ continuum ($b_\mathrm{corr}$) as a function of the projected baseline length ($B$). Symbols are color-coded for observation date (color-coding is the same as in Fig.~\ref{fig:model}).} 
 \label{fig:B_vs_r}
        \end{figure}

Using interferometric observations, it is possible to narrow down the spatial regions responsible for the silicate emission \citep[e.g.,][]{Boekel_silicate}. The mid-infrared measurements of \cite{2016Millan-Gabet} already suggested a difference between the correlated and total spectra in 2010. While DG Tau did not show emission in the total spectrum, it showed indication of silicate in absorption in the correlated spectrum. Our VLTI/MIDI observations strengthen this notion. The correlated and total spectra of DG Tau have dramatically different shape. In the following we attempt to decompose the correlated spectra (Sect. \ref{sec:correlated}; dominated by the inner few au region of the circumstellar disk) and the uncorrelated spectra (Sect. \ref{sec:uncorr}; the difference between the total and correlated spectra, which thus describes the outer, large-scale part of the disk) of DG Tau into a linear continuum and the silicate feature. Using the resulting parameter values we then study the changes in the silicate feature in different epochs and at different resolutions.






\subsubsection{Spectra of the inner disk} \label{sec:correlated}



The correlated spectra, shown in Fig. \ref{fig:corr_matrix}, are dominated by the inner, compact region of the disk, whose size is determined by the resolution of the interferometer for the given baseline (see Table~\ref{table:obs}). All of them are very smooth. The main trend is a decrease up to $\sim$\,$9.5~\mu$m and a flat or increasing part at longer wavelengths. This shape suggests that we see the silicate feature in absorption here, and the profile and the wavelength of the minimum indicates amorphous grains.
   
     \begin{table}
\caption{Results of the fit to the correlated spectra.}             
\label{table:corr_fit}      
\centering                          
\begin{tabular}{l l l l l}        
\toprule \toprule  
Epoch & $B$ & $b_\mathrm{corr}$  & $a_\mathrm{corr}$  & $c_\mathrm{corr}$ \\ 
 & (m) & (Jy) & (Jy/$\mu$m) & (Jy) \\
\midrule                   
   2011 Oct & $33$ & $3.2 \pm 0.2$ & $0.25 \pm 0.02$ & $0.29 \pm 0.02$ \\ 
    & $76$ & $2.6 \pm 0.2$ & $0.07 \pm 0.01$ & $0.41 \pm 0.03$ \\
    \midrule
   2011 Dec & 56 & $3.3 \pm 0.2$ & $0.12 \pm 0.01$ & $0.42 \pm 0.03$ \\ 
   & 37 & $4.0 \pm 0.2$ & $0.24 \pm 0.01$ & $0.36 \pm 0.02$ \\
   \midrule
   2012 Feb & 44 & $3.3\pm 0.2$ & $0.18 \pm 0.01$ & $0.43 \pm 0.08$ \\
    & 83 & $2.2\pm 0.2$ & $0.031\pm 0.004$ & $0.20 \pm 0.02$ \\
   2012 Nov & 64 & $3.3 \pm 0.2$ & $0.09\pm0.01$ & $0.48 \pm 0.03$ \\
   \midrule
   2014 Jan & 88 & $2.3 \pm 0.1$ & $-0.006 \pm 0.002$ & $0.30 \pm 0.04$ \\
    & 89 & $2.2\pm 0.2$ & $-0.05 \pm 0.02$& $0.27 \pm 0.08$ \\
\bottomrule                                  
\end{tabular}
\end{table}

To describe the correlated spectra in Fig. \ref{fig:corr_matrix}, we model them as a combination of a linear continuum and an amorphous silicate absorption. The spectra, $F_{\mathrm{corr}}(\lambda)$ were fitted with the following formula:
\begin{equation}
F_{\mathrm{corr}}(\lambda) = a_\mathrm{corr} \cdot \left(\lambda - 10.7\,\mu\mathrm{m}\right) + b_\mathrm{corr} - c_\mathrm{corr} \cdot G(\lambda),
\end{equation}
where $a_\mathrm{corr}$, $b_\mathrm{corr}$, and $c_\mathrm{corr}$ are wavelength independent factors and $G(\lambda)$ is the normalized spectrum of interstellar amorphous silicate grains measured by \cite{Kemper_silicate} towards the Galactic Center. 
The correlated spectra are fitted in the wavelength range between $8~\mu$m and $12.7~\mu$m. We sampled a grid of the parameter values of $a_\mathrm{corr}$, $b_\mathrm{corr}$, and $c_\mathrm{corr}$. The fits were compared using the obtained $\chi^2$ values. We note that the parameters were fitted simultaneously, making it possible to determine a reliable continuum level, even though the broad absorption feature would not allow to constrain the continuum alone. The best-fit parameter values corresponding to the lowest $\chi^2$ are listed in Table \ref{table:corr_fit}. In Appendix \ref{app:err} we describe how uncertainties of the parameters were calculated. In Fig. \ref{fig:corr_matrix} the best fitting $F_{\mathrm{corr}}(\lambda)$ curves are displayed. Apart from the shortest investigated wavelengths ($8-9~\mu$m), the resulting curves are good representations of the measurements. 


The derived absolute flux density levels at $10.7$\,$\mu$m ($b_\mathrm{corr}$ in Table \ref{table:corr_fit}) are mainly determined by the projected baseline length as shown in Fig \ref{fig:model}. The slope of the linear continuum ($a_\mathrm{corr}$) monotonically decreases with increasing projected baseline length. This is expected, since with better resolution (longer projected baseline lengths) we observe the more compact regions, which are dominated by the hotter dust emitting at shorter wavelengths. At the longest baselines ($83$\,m in 2012 February and $88$\,m and $89$\,m in 2014 January), the derived continuum is almost completely flat. Since both $a_\mathrm{corr}$ and $b_\mathrm{corr}$ have a dependence on $B$, the two parameters correlate: their correlation coefficient is $0.87$.

The average value of the amplitude of the silicate absorption is $\langle c_\mathrm{corr} \rangle = 0.35$\,Jy with a standard deviation of $\sigma_{c,\mathrm{corr}} = 0.09$\,Jy. In Fig. \ref{fig:B_vs_r}, we show the amplitude of the absorption normalized for the $10.7$\,$\mu$m continuum ($c_\mathrm{corr}/b_\mathrm{corr}$) as a function of the projected baseline length ($B$). There is a clear  trend of increasing $c_\mathrm{corr}/b_\mathrm{corr}$ with $B$, which might indicate that the absorption mostly originates from a confined region in the inner part of the dusty disk. It is interesting to note, that $c_\mathrm{corr}/b_\mathrm{corr}$ values at $B > 80$~m do not fit in this trend. This may be a hint that the absorption comes from a ring-shaped region with inner and outer radii of $\sim$\,$1$~au and $\sim$\,$3$~au, respectively (these values refer to the resolution of the interferometer at the corresponding baselines). 
The presence of a general trend in Fig.~\ref{fig:B_vs_r} suggests that the main variations in the normalized absorption are related to the different spatial resolutions. Thus there is no 
indication of temporal variability within the inner regions. 




\subsubsection{Spectra of the outer disk} \label{sec:uncorr} 

\begin{figure*}
        \centering
        \includegraphics[angle=-90, width=17cm]{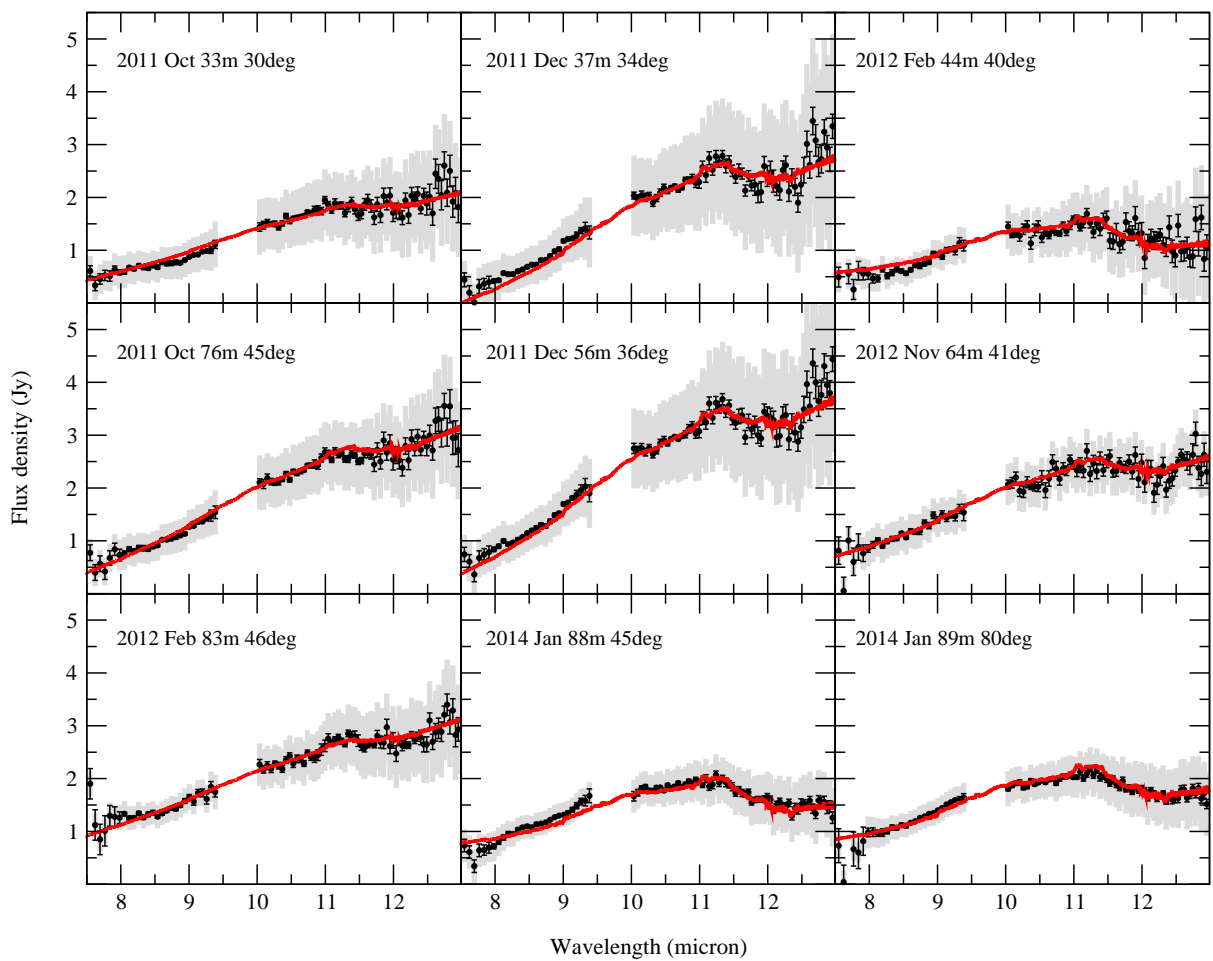}
        \caption{Uncorrelated spectra (black symbols) of DG Tau at different projected baselines and at different epochs. The part affected by the telluric ozone feature, between $9.4\,\mu$m and $10\,\mu$m is not shown. Red line represents the best-fit curve using the silicate spectrum of Hale-Bopp (Eq. \ref{HB}). The rows are categorized by approximate size of the observed baseline (top: short baseline, middle: intermediate baseline, bottom: long baseline). Gray shading indicates the level of total calibration uncertainties, while black error bars represent the random point-to-point uncertainties (see Appendix \ref{app:err} for more details on error calculation). 
\label{fig:uncorr_matrix}}
\end{figure*}

\begin{table}
\caption{Results of the fit to the uncorrelated spectra.}             
\label{table:uncorr_fit}      
\centering                          
\begin{tabular}{l l l l l}        
\toprule \toprule                 
Epoch & $B$ & $b_\mathrm{uncorr}$  & $a_\mathrm{uncorr}$  & $c_\mathrm{uncorr}$  \\   
 & (m) & (Jy) & (Jy/$\mu$m) & (Jy) \\
\midrule                      
2011 Oct & $33$ & $1.4 \pm 0.6$ & $0.3 \pm 0.1$ & $0.3 \pm 0.1$ \\
         & $76$ & $2.0 \pm 0.6$ & $0.5 \pm 0.1$ & $0.5 \pm 0.1$  \\ 
\midrule
2011 Dec & $56$ & $2.3 \pm 1.0$ & $0.6 \pm 0.2$ & $0.9 \pm 0.2$ \\
         & $37$ & $1.6 \pm 1.0$ & $0.5 \pm 0.2$ & $0.8 \pm 0.2$  \\ 
\midrule
2012 Feb & $44$ & $0.9 \pm 0.6$ & $0.1 \pm 0.1$ & $0.7 \pm 0.2$  \\ 
         & $83$ & $2.2 \pm 0.5$ & $0.4 \pm 0.1$ & $0.3 \pm 0.1$  \\
\midrule
2012 Nov & $64$ & $1.8 \pm 0.4$ & $0.34 \pm 0.04$ & $0.6 \pm 0.1$  \\ 
\midrule
2014 Jan & $88$ & $1.2 \pm 0.4$ & $0.1 \pm 0.1$ & $0.8 \pm 0.1$  \\
         & $89$ & $1.4 \pm 0.4$ & $0.2 \pm 0.1$ & $0.8 \pm 0.1$  \\ 
\bottomrule                              
\end{tabular}
\end{table}

The uncorrelated spectrum is the emission component that is spatially resolved out by the interferometer, thus does not appear in the correlated spectrum. It can be computed by subtracting the correlated spectrum from the total spectrum. In the case of our VLTI/MIDI observations uncorrelated spectra are dominated by the outer parts of the $10~\mu\mathrm{m}$ emitting region of the circumstellar disk, outside a radius of $1-3$ au (depending on the resolution, see Table \ref{table:obs}). In contrast to the previous subsection, the uncorrelated spectra show no absorption; rather we see an emission feature (see Fig. \ref{fig:uncorr_matrix}), similar to the observations of  \cite{Bary_spitzer}, for example. We note that \cite{Bary_spitzer} did not have spatially resolved observations, that is, they studied total spectra.

\begin{figure}
\resizebox{\hsize}{!}{\includegraphics[ clip=]{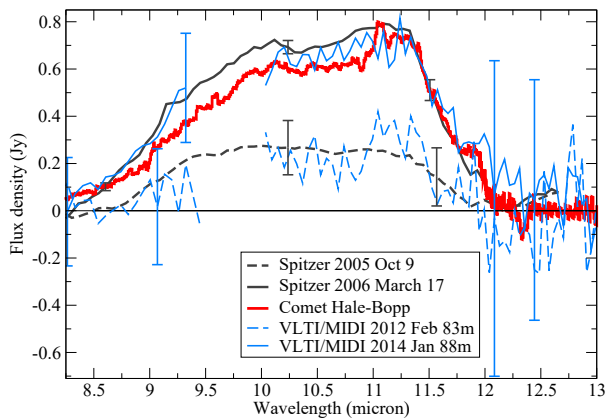}}\caption{Emission feature in the spectra of DG Tau. Dashed and solid blue lines show the continuum subtracted uncorrelated spectra of DG Tau in two epochs. Dashed and solid gray lines show the continuum subtracted spectra measured by {\it Spitzer} in 2005 October 9 and 2006 March 17, respectively \citep{Bary_spitzer}, corrected for an average absorption (see text for details). The red line represents the emission feature in the spectrum of comet Hale-Bopp, scaled to match the solid blue curve. \label{fig:uncorr_subtr}}
\end{figure}

In order to describe the shape of the uncorrelated spectra, we have to choose a spectral template for the silicate emission feature. \citet{Bary_spitzer} obtained high-resolution Spitzer spectra of the silicate feature, indicating the presence of crystalline forsterite. Previous studies \citep[e.g.,][]{Abraham2009} showed that the silicate feature in the spectrum of comets can be very similar to that of protoplanetary disks. Therefore we chose the continuum-subtracted, normalized mid-infared spectrum of comet Hale-Bopp \citep[][and see red curve in Fig. \ref{fig:uncorr_subtr}]{Hale-Bopp} as a template to fit the silicate emission feature. This spectrum also shows the presence of crystalline silicate grains.  Since the general shape of the Hale-Bopp spectrum matches well the spectrum of DG Tau, it was unnecessary to introduce any correction for possible temperature differences between the comet and the disk. We model the spectra by the following formula:
%
%
\begin{equation} \label{HB}
F_{\mathrm{uncorr}}(\lambda) = a_\mathrm{uncorr} \cdot (\lambda - 10.7\,\mu\mathrm{m}) + b_\mathrm{uncorr} + c_\mathrm{uncorr} \cdot H(\lambda),\end{equation}
with parameters $a_\mathrm{uncorr}$ (slope) and $b_\mathrm{uncorr}$ ($10.7\,\mu\mathrm{m}$ flux density), and the silicate emission template ($H(\lambda)$) from the spectrum of comet Hale-Bopp with an amplitude $c_\mathrm{uncorr}$.

The derivation of the best-fit parameters and their uncertainties was done in the same way as for the correlated spectra (see Sect. \ref{sec:correlated}). We note that the uncertainties of the measured points are much larger for the uncorrelated spectra than for the correlated spectra. This is caused by the larger uncertainties of the total spectra due to the high systematic background residuals (see Sect.~\ref{sec:calqual}). In the case of the correlated spectra the background noise almost completely cancels out since it is not coherent \citep{midi_reduc1}. Hence, the correlated spectra are only limited by the statistical photon noise. Additionally, the wavelength range affected by the atmospheric ozone feature (between $9.4\,\mu$m and $10\,\mu$m) was disregarded in the fitting process. The resulting best-fit parameter values are given in Table \ref{table:uncorr_fit}. The best fit curves are overplotted on the data in Fig. \ref{fig:uncorr_matrix}. The fits reproduce the measured data points reasonably well, and the silicate emission feature is clearly detected at least at the $3\sigma$ level in all epochs. It is quite remarkable that the spectrum of a comet is very similar to the spectrum of a protoplanetary disk.

The average value of the amplitude of the silicate emission is $\langle c_\mathrm{uncorr} \rangle = 0.63$~Jy with a standard deviation of $\sigma_{c,\mathrm{uncorr}} = 0.22$~Jy, which is larger than the value for the absorption in the correlated spectra ($\sigma_{c,\mathrm{corr}}$) by a factor of $2.5$, which means more significant variability. If we take the relative scatter (i.e., $\sigma_c / \langle c \rangle$), then we still see that the variations in the emission are about one third higher than the variations in the absorption. 

Changes in $c_\mathrm{uncorr}$ can also be traced in Fig.~\ref{fig:uncorr_subtr}, where we show the continuum subtracted emission features from our MIDI data (blue curves), and from earlier {\it Spitzer} measurements of \citet{Bary_spitzer} (gray curves). In order to be able to compare the {\it Spitzer} and MIDI spectra, we only took into account the common wavelength range ($8-13~\mu$m). Additionally, the continua in the {\it Spitzer} spectra were calculated in the same way as in the MIDI uncorrelated spectra by fitting them with Eq. \ref{HB}.  \citep[We note that in general the wider wavelength range where {\it Spitzer} operated allows a more precise continuum fitting, as was shown in][]{Bary_spitzer}. {\it Spitzer} observed the total spectra, so to be able to directly compare these data with the uncorrelated MIDI spectra, a correction for the inner disk absorption is also needed. Thus we subtracted an absorption profile with an amplitude of $0.25$~Jy from the {\it Spitzer} spectra. The amplitude was chosen to match with the MIDI observations with the corresponding baselines. In both the {\it Spitzer} and MIDI datasets a large and small silicate feature is shown. The amplitudes and shapes of the silicate features are relatively similar. We also show in Fig.~\ref{fig:uncorr_subtr} the spectrum derived from the comet Hale-Bopp multiplied by $0.8$, corresponding to best-fit value obtained for the January MIDI uncorrelated spectrum (red line). 




  \begin{figure}
\resizebox{\hsize}{!}{\includegraphics{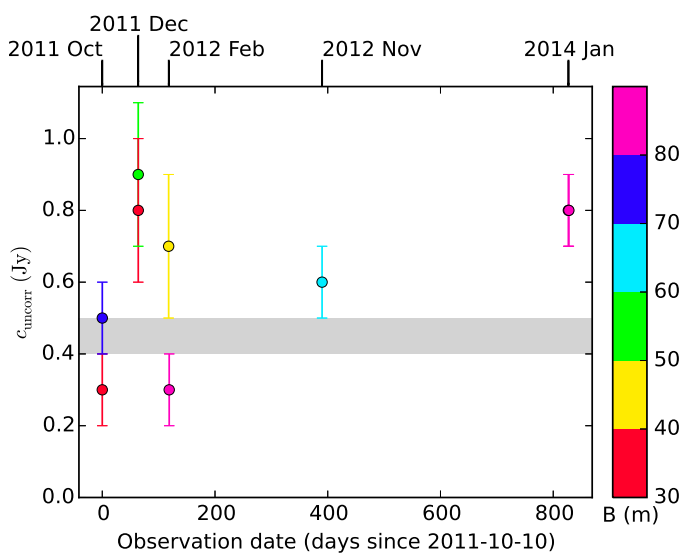}}
                \caption{Amplitude of the silicate emission feature ($c_\mathrm{uncorr}$) plotted as a function of observation date. Symbols are color-coded for the projected baseline ($B$). Notice the large temporal variations of $c_\mathrm{uncorr}$. The gray stripe indicates the range of the emission feature strength obtained from radiative transfer modeling, see Sect. \ref{sec:emission_radmodel} for details.} 
\label{fig:mjd_vs_c}
        \end{figure}
        
  \begin{figure}
\resizebox{\hsize}{!}{\includegraphics{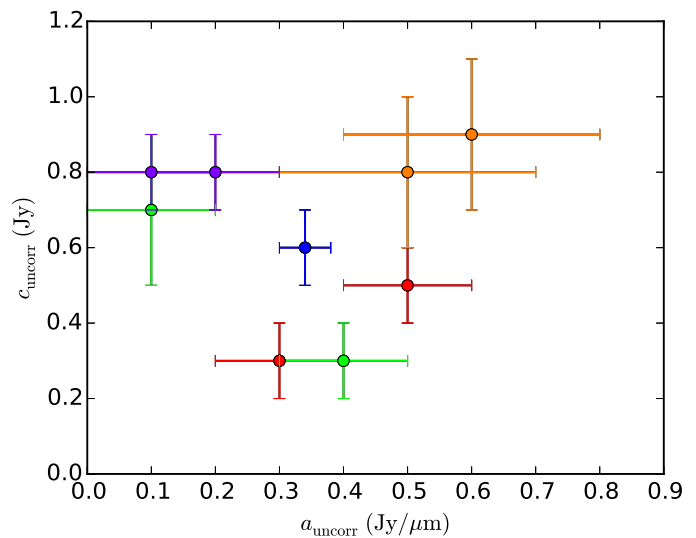}}
                \caption{Amplitude of the silicate emission feature ($c_\mathrm{uncorr}$) plotted as a function of the uncorrelated continuum slope ($a_\mathrm{uncorr}$). Symbols are color-coded for observation date (color-coding is the same as in Fig.~\ref{fig:model}). 
} 
\label{fig:a_vs_c}
        \end{figure}
    
To disentangle temporal and spatial variability in interferometric observations one has to consider observations taken at similar projected baseline lengths at different epochs, and observations taken at different projected baseline lengths at the same epoch. In Fig.~\ref{fig:mjd_vs_c} we plot the amplitude of the silicate emission feature ($c_\mathrm{uncorr}$) as a function of time, color-coded for projected baseline length. Comparing the observations taken at similar angular resolutions in 2012 February and 2014 January at a projected baseline length of $\sim 85$\,m (purple points), and in 2011 October and December at a projected baseline length of $\sim 33$\,m (red points) we find significant variability in the amplitude of the emission feature over time. At the shortest baselines, $c_\mathrm{uncorr}$ had the lowest value in 2011 October, but it almost tripled two months later. At the longest baselines we see the same amount of change from 2012 February to 2014 January.  
To deduce the spatial variability, we can compare observations taken at the same epochs with different baseline configurations. According to Fig.~\ref{fig:mjd_vs_c}, we do not see any trend in the silicate emission feature as a function of the baseline length. We note however, that the uncorrelated spectra are much less sensitive on the baseline than the correlated spectra, because the uncorrelated emission comes from a much larger area.

\citet{Bary_spitzer} presented a figure of the silicate emission amplitude as a function of the  $\left[6.0 \right]-\left[13.5 \right]$ $\mu$m color index, based on their Spitzer data. They found a weak correlation between the two quantities. Instead of this color index we can use the continuum slope $a_\mathrm{uncorr}$ to explore this relation. In Fig.~\ref{fig:a_vs_c} we plot $c_\mathrm{uncorr}$ versus the mid-infrared spectral slope $a_\mathrm{uncorr}$, from our uncorrelated MIDI spectra. The correlation coefficient between $a_\mathrm{uncorr}$ and $c_\mathrm{uncorr}$ is $\rho_{a,c,\mathrm{uncorr}} = -0.04$, thus we find no correlation between the two parameters. 

Previously, in the correlated spectra we have seen a correlation between the continuum slope ($a_\mathrm{corr}$) and the $10.7\,\mu\mathrm{m}$ flux density ($b_\mathrm{corr}$). This relation is a consequence of the different spatial scales probed by the interferometer: with a longer baseline, the correlated spectrum comes from a smaller region, thus having less absolute flux densities. At the same time, we see the hotter inner parts of the disk having bluer color and less steep spectral slope. Interestingly, we observe a similar relation in the uncorrelated spectra between $a_\mathrm{uncorr}$ and $b_\mathrm{uncorr}$, with $\rho_{a,b,\mathrm{uncorr}} = 0.86$ (the same value as in the case of $a_\mathrm{corr}$ and $b_\mathrm{corr}$). This means that when the outer disk gets brighter, its spectral slope gets steeper (corresponding to a lower average temperature). But neither $a_\mathrm{uncorr}$ nor $b_\mathrm{uncorr}$ exhibit significant dependence on the baseline length. Thus we attribute the scatter in these parameters to temporal variations that cause changes in the continuum in a way that preserves the relation between $a_\mathrm{uncorr}$ and $b_\mathrm{uncorr}$.

\section{Discussion}
\label{sec:discuss}

\subsection{Origin of the mid-infrared variability}

\label{sec:disc_var}




While an increasing number of studies demonstrated that mid-infrared variability is common among young stars \citep[e.g.,][]{kospal_atlas,Bary_spitzer}, DG Tau exhibits unusually large amplitudes up to a factor of two in the $10~\mu$m silicate feature. Moreover, the feature may even turn into absorption, which is very rarely observed in other young stars. \cite{Bary_spitzer} developed a two-temperature two-slab model to explain the variability of the silicate emission feature, based on the method of \citet{Sargent2009_bary}. This model includes two emitting layers on the disk surface, with the top, intervening layer being cooler, enabling self-absorption. The authors were able to reproduce their multi-epoch data set with this model by fitting the temperatures and dust masses. An important finding is  
that the intervening cooler dust layer has the same composition as the emitting region, therefore the variations do not indicate any change in the dust composition. \citet{Bary_spitzer} proposed three scenarios to explain the observed silicate variability on month- and week-long timescales: a) variable disk-shadowing by a warped puffed-up inner rim, b) turbulent mixing, and c) disk winds, capable of lifting dusty material above the disk.

One of the main results of our MIDI observations is that the silicate emission and absorption are detected in spatially distinct regions in the disk. 
Thus, the variability observed in the total spectrum of DG Tau can be explained as the combination of an almost constant absorption towards the inner disk and a variable silicate emission from the outer disk outside $1-3$~au. If the emission feature completely fills up the absorption dip, no feature is seen in the total spectrum  (e.g., in the TIMMI2 observation in 2002, see Fig. \ref{fig:timmi2}), while stronger emission gives rise to the silicate feature usually observed. 

The other important finding of our analysis is that the shape of the absorption feature in the correlated spectra is characteristic of amorphous grains, but the emission feature in the uncorrelated spectra indicates the presence of crystalline grains.
This is an unexpected result, as it seems to be in contradiction with the usual picture of protoplanetary disks, where thermal annealing causes the crystallization of initially amorphous grains \citep{Colangeli2003} in the inner disk, near the sublimation radius, while in the outer disk the grains are largely left unprocessed \citep{Boekel_silicate}. In the case of DG Tau, however, we see that the highly processed dust lies in the outer regions of the mid-infrared emitting disk (outside $1$ au). 
In the following we attempt to explain the spatially distinct silicate emission and absorption features.

\subsection{Radiative transfer modeling} \label{radmodel}

\begin{table}
\caption{Parameters used in our radiative transfer model. Those in italics were kept fixed during the modeling.
}
\label{tab:radmc}
\centering
\begin{tabular}{l p{4.6cm} r l}
\toprule \toprule
\multicolumn{4}{l}{\textbf{System Parameters}}\\
& \emph{Distance} ($d$)                               & 140    &pc\\
& \em{Inclination ($i$)}                    &28      &$^{\circ}$\\
\multicolumn{4}{l}{\textbf{Stellar Parameters}}\\
& \emph{Temperature} ($T_*$)       &              4300    &K\\
& \emph{Luminosity}  ($L_*$)                           &1.0     & L$_\sun$ \\
&\emph{Mass}  ($M_*$)                           &1.0     & M$_\sun$ \\
&\emph{Visual extinction} ($A_\mathrm{V}$)                    &  1.6      &\\
&\emph{Selective extinction coefficient} ($R_\mathrm{V}$)                    &  3.1      &\\
\multicolumn{4}{l}{\textbf{Gaseous Accretion Disk}}\\
&{Inner radius} ($R_\mathrm{in}^\mathrm{g}$)    & 0.01      & au\\
&{Outer radius} ($R_\mathrm{out}^\mathrm{g}$)   &0.07  &au\\
&{Accretion rate} ($\dot{M}$)   &$10^{-7}$  &$\mathrm{M}_\sun \mathrm{yr}^{-1}$\\
\multicolumn{4}{l}{\textbf{Dusty Disk}}\\
&{Inner radius} ($R_\mathrm{in}^\mathrm{d}$)    & 0.07      & au\\
&{Outer radius} ($R_\mathrm{out}^\mathrm{d}$)   &100  &au\\
&Inner scale height ($h_\mathrm{in}^\mathrm{d}$)               & 0.04     &\\
&Power-law index for scale height ($p_h$)               & 0.228    &\\
&{Power-law index for surface density ($p_\Sigma$)}               & 0    &\\
&Mass ($M_\mathrm{disk}$)               & $3.4\times10^{-2}$     & M$_\sun$\\
&\emph{Gas-to-dust mass ratio} & $100$     & \\
\multicolumn{4}{l}{\textbf{Dust Grain Size Distribution}}\\
&\emph{Diameter range} ($a$) & $[10^{-2},~10^{3}]$ & $\mu$m \\
&\emph{Power-law index} ($p$) & $-3.5$ &  \\
&\emph{Dust species:} & &\\
& \hspace{0.3cm}\emph{amorphous carbon} & 50 & $\%$\\
& \hspace{0.3cm}\emph{astronomical silicate} & 50 & $\%$\\
\multicolumn{4}{l}{\textbf{Cloud Component}}\\
&Inner radius ($R_\mathrm{in}^\mathrm{c}$)    & 5      & au\\
&Outer radius ($R_\mathrm{out}^\mathrm{c}$)   &7.5  &au\\
&Center height ($h_\mathrm{center}^\mathrm{c}$)               & 0.6     &\\   
&Vertical size  ($\Delta h_\mathrm{cloud}$)               & 0.6     &\\ 
&Mass ($M_\mathrm{cloud}$)               & $3.7\times10^{-7}$,      & M$_\sun$\\
&               & $7.4\times10^{-7}$     & M$_\sun$\\
&\emph{Gas-to-dust mass ratio} & $100$     & \\

\bottomrule
\end{tabular}

\label{tab:para}
\end{table}

To evaluate the different scenarios we use the radiative transfer code RADMC-3D\footnote{\scriptsize \url{http://www.ita.uni-heidelberg.de/~dullemond/software/radmc-3d/}} \citep{Dullemond2012}. We assume a flared, axisymmetric circumstellar disk geometry, where the inner radius is set at the sublimation radius. DG Tau has a considerable accretion rate, thus we include a gaseous accretion disk inside the dust sublimation radius. We also take into consideration  the accretion heating in the mid-plane of the dusty disk, thus it is heated by both the accretion and the illumination from the central engine.
For computing the spectral energy distribution (SED) of the inner gaseous disk, we adopt the Shakura-Sunyaev disk description \citep{Shakura1973}, where a geometrically thin, optically thick disk
emits black-body radiation with a surface effective temperature,
\begin{equation}
T^4\left(r\right)=
\frac{3G M_* \dot{M}}
{8\pi\sigma r^3}
[1-\left(R_\mathrm{in}^\mathrm{g}/r\right)^{1/2}],
\end{equation}
where $G$ is the gravitational constant, $\sigma$ is
the Stefan-Boltzmann constant, $M_*$ is the stellar mass, $\dot{M}$ is the accretion rate and $R_\mathrm{in}^\mathrm{g}$ is the inner radius of the gaseous accretion disk. For the dusty disk we assume a power-law radial distribution for its surface density $\Sigma$,
\begin{equation}
\Sigma\left(r\right) = \Sigma_\mathrm{in}\left(\frac{r}{R_\mathrm{in}^\mathrm{d}}\right)^{p_\Sigma}
,~
R_\mathrm{in}^\mathrm{d} < r < R_\mathrm{out}^\mathrm{d},
\end{equation}
where $\Sigma_\mathrm{in}$ is the surface density at the inner radius $R_\mathrm{in}^\mathrm{d}$, $p_\Sigma$ is the power-law exponent, and  $R_\mathrm{out}^\mathrm{d}$ is the outer radius of the dusty disk. We assume the vertical distribution to be a Gaussian function, and the resulting density structure of the disk is
\begin{equation}
\rho \left(r,z\right) = \Sigma\left(r\right)\frac{1}{\sqrt{2\pi}H} \exp \left(-\frac{z^2}{2 H^2} \right),
\end{equation}
where $\rho(r,z)$ denotes the dust density as a function of $r$ and the height $z$ above the mid-plane, and $H$ is the scale height. We assume that the dependence of $H$ on $r$ is also a power-law,  
\begin{equation}
h\left(r\right) \equiv \frac{H\left(r\right)}{r} = h_\mathrm{in}^\mathrm{d}\left(\frac{r}{R_\mathrm{in}^\mathrm{d}}\right)^{p_h},
\end{equation}
where $h\left(r\right)$ is the dimensionless scale height, $h_\mathrm{in}^\mathrm{d}$ is the dimensionless scale height at the inner radius, and $p_h$ is the power-law index.
To account for the accretion heating within the dusty disk, we again adopt the Shakura-Sunyaev disk description,
where the surface power density $P_\mathrm{acc}$ from accretion heating is
\begin{equation}
P_\mathrm{acc} (r)=
\frac{3G M_* \dot{M}}
{4\pi r^3}
[1-\left(R_\mathrm{in}^\mathrm{g}/r\right)^{1/2}]
,~
R_\mathrm{in}^\mathrm{d} < r < R_\mathrm{out}^\mathrm{d}
.
\end{equation}
Heat sources are added to the mid-plane of the dusty disk according to the above equation.
We assume $\dot{M}$ to be a constant throughout the whole disk range, including the inner gaseous disk and the dusty disk.
The refractive index for the dust species is taken from \citet[for amorphous carbon]{Jager1998carbon} and \citet[for astronomical silicate]{Draine1984}.

\begin{figure}
\resizebox{\hsize}{!}{\includegraphics{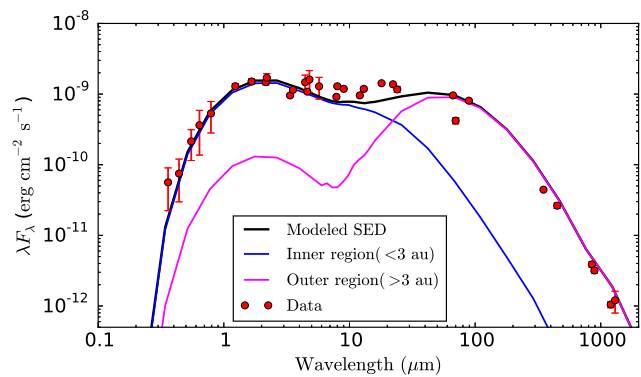}}
\caption{SED (black line) from the radiative transfer model, together with the observed
SED of DG~Tau (red circles).
We also plot the flux contributions from different radial regions: the inner region including the star and the inner disk ($<3$~au, blue) and the outer disk ($>3$~au, purple). We note that in the optical regime time-averaged data points are shown and their error bars indicate temporal variability. The data points are from the literature \citep{1992A&A...258..217E,2000yCat.1261....0K,2004AAS...205.4815Z,2008A&A...488..401R,2012yCat.1322....0Z,2014yCat.1327....0N,2015AJ....150..101Z,1978A&AS...34..477M,2006PASP..118.1666D,2016yCat.2336....0H,1988cels.book.....H,2012ApJS..203...21A,2014yCat....1.2023S,2011ApJ...731....8K,2010ApJS..187..149A,2001A&A...365...90F,2007A&A...461..183G,1996A&AS..117..335A,2002A&A...386..492R,2005A&A...431..773R,2010AJ....140.1214C,2MASS_cat,Bary_spitzer,kospal_atlas,2006ApJ...647.1180L,2009ApJ...696L..84C,2013ApJ...769...21S,WISE_cat,AKARI_cat1,AKARI_cat2,SCUBA_flux,2013ApJ...771..129A,IRAM_flux,VLA_flux}.}
\label{fig:sed}
\end{figure}

All stellar and system parameters  used in our radiative transfer model are given in Table~\ref{tab:radmc}. To account for the extinction along the line-of-sight, we applied the reddening curve of \citet{CCM1989}, and adopted an optical extinction value $A_\mathrm{V} = 1.6$~mag by \citet{Furlan2009}. In Fig.~\ref{fig:sed} we plot the SED of DG Tau along with the curves from the RADMC-3D model. SEDs of different disk regions were calculated using masks. The model broadly agrees with the observations in the optical, near infrared, and far infrared. The mid-infrared wavelength range is addressed in Sect. \ref{sec:emission_radmodel}. 

\subsection{Absorption in the correlated spectra}

Here we present four possible scenarios for explaining the observed absorption feature in the inner disk spectra.

\subsubsection{Cold absorbing envelope in the line of sight}

The absorption feature in the correlated spectrum could potentially be caused by a cold, steady dust cloud, or envelope, in the line of sight between the disk and the observer. Our MIDI observations reveal that this intervening cloud is comprised of pristine, amorphous, interstellar-like particles. The question is, however, whether the extinction value needed to produce the observed $10~\mu$m absorption feature depth would be consistent with the optical extinction measured towards the star. We performed a simple calculation, assuming the extinction law by \cite{mathis_ext} and adopting the higher number of the two $A_\mathrm{V}$ values found in the literature \citep[$A_\mathrm{V} = 1.6$~mag, $A_\mathrm{V} = 3.2$~mag][respectively]{Furlan2009, Rebull2010}. After dereddening the correlated spectra, the depth of the absorption decreases by $50\%$, but it still remains  visible. We would need $A_\mathrm{V} = 6-7$~mag to fill the silicate absorption entirely. If we accept this value for reddening, then the star should be intrinsically brighter and bluer than has been reported in the literature. Stellar SED fits usually show degeneracy between the effective temperature of the photosphere and $A_\mathrm{V}$. However, the spectral type (K6$-$K8) and effective temperature of DG Tau are well constrained from optical spectral line measurements \citep{Herczeg2014}, reducing the possibility for an $A_\mathrm{V}$ much larger than $3$~mag.

For an independent check of this result, we implemented an external absorbing structure in our radiative transfer model. We run models with a spherical envelope component placed at $R>100~\mathrm{au}$, assuming pristine dust similar to the interstellar medium, with the grain size distribution $a=[10^{-2},~10^{3}]~\mu$m, with a power-law index $p=-3.5$. The modeling showed that it is impossible to create an absorption feature with the observed strength without causing optical extinction much stronger than the observed value. This difficulty could only be circumvented if the dust has larger grain size and therefore a flatter opacity curve. However, it does not seem plausible for evolved dust grains to exist at such a distance from the star. Thus, we conclude that an external cold intervening dust layer can be responsible for part ($\sim 50\%$) of the observed absorption feature in the correlated spectra.

In Sect.~\ref{sec:correlated} we have shown that the normalized absorption strength ($c_\mathrm{corr}/b_\mathrm{corr}$) increases with baseline, which means that $c_\mathrm{corr}/b_\mathrm{corr}$ decreases with radius in the inner disk (see Fig.~\ref{fig:B_vs_r}). To interpret the change in the strength of the absorption feature we have two options. Either a) we have a combination of a homogeneous envelope and a spatially varying emission feature in the disk.  
The emission would gradually fill up the absorption dip as we move farther away from the star and even turn the spectrum into emission in the outer disk, which is observed in the uncorrelated spectra. This might be due to the higher fraction of larger grains in the inner disk, indicating ongoing dust processing, or b) we assume an absorbing structure confined to the innermost regions of the disk ($<1-3$~au). This scenario is further discussed in the following subsections. 


\subsubsection{Absorption layer due to accretion heating}

The absorption feature, partly or fully, could be caused by a colder upper layer on the inner dusty disk, which absorbs radiation from the warmer disk material in the mid-plane. In order to be consistent with our MIDI observations, we assume that this upper layer includes small amorphous silicate particles.
This kind of ``inverse'' temperature gradient can be caused by a sufficiently strong accretion, which can heat up the disk from the mid-plane close to the star. The accretion rate was estimated for DG Tau with several methods in various epochs; the measurements range from $9.6 \times 10^{-8}~\mathrm{M}_\sun \mathrm{yr}^{-1}$ to $2 \times 10^{-6}~\mathrm{M}_\sun \mathrm{yr}^{-1}$, with a median value of $3.5 \times 10^{-7}~\mathrm{M}_\sun \mathrm{yr}^{-1}$ \citep{Hartigan1995,muzerolle_brgamma,gullbring_accretion,WhiteHillenbrand2004,ca_accretion,brgamma_accretion,accretionrates}. These values are at least three times higher than the average mass accretion rate of T Tauri stars \citep[$3 \times 10^{-8}~\mathrm{M}_\sun \mathrm{yr}^{-1}$, ][]{IMTTS_accretion}.


Our model described in Sect. \ref{radmodel} uses an accretion rate of $10^{-7}~\mathrm{M}_\sun \mathrm{yr}^{-1}$, which cannot reproduce the silicate absorption feature in the inner disk (see blue curve in Fig.~\ref{fig:sed}). Therefore, we tested radiative transfer models with even higher accretion rates. Our modeling shows that, in order to reproduce the silicate absorption feature in the correlated flux, we need the unphysical assumption
that the higher accretion rate does not lead to enhanced irradiation of the disk surface. Even in this case the accretion rate must be increased to $\dot{M} = 10^{-6} ~\mathrm{M}_\sun \mathrm{yr}^{-1}$ in order to get a silicate absorption. Although this value is within the observed range, the heating from such strong accretion would also increase the flux densities in the wavelength range $3-7~\mu\mathrm{m}$ to values higher than observed. Additionally, the higher accretion rate will also cause absorption lines in optical and NIR wavebands, which is not seen in the observational data of DG Tau \citep[e.g.,][]{DGTau_optical_spectrum,DGTau_jet2016}. Therefore, our modeling does not favor the high accretion rate as the sole cause of the absorption feature.




\subsubsection{Temperature inversion on the disk surface}

\citet{Vinkovic2006} presented an analytical study of radiative transfer on the surface of externally heated dust clouds. The paper specifically discusses a scenario where larger grains ($> 1 \mu$m) in the inner disk are heated such that the maximum temperature occurs at the $\tau_\nu \sim 1$ surface. The author finds that when a grain size distribution is included, small grains are sublimated away from the very surface of the inner disk edge, while large grains can survive there. In the resulting surface region ($0 < \tau_\nu < 1$), dominated by large grains, the temperature decreases outwards, thus it is capable of producing a silicate absorption profile. Viscous heating of the inner disk due to accretion would presumably strengthen this effect. Given that the MIDI observations seem to support stronger silicate absorption at higher resolutions, this may be a plausible mechanism to explain absorption at the inner rim of the disk. However, the absorption profile in our MIDI data is produced by ISM-like amorphous dust grains. Large evolved dust grains, needed for this scenario, would possibly show flatter silicate absorption profile, observed in systems with progressed grain growth \citep{Przygodda2003}.

\subsubsection{Misaligned inner disk}

Here we consider an additional scenario with a special disk geometry to explain the silicate absorption. It is known that in some YSOs the inner and outer disks are misaligned \citep[see e.g.,][for the case of LkCa 15 and references therein]{Thalmann2015}. If the inner disk of DG Tau is seen almost edge on, we would expect to see silicate absorption due to the high column density and outwards decreasing temperature in the line of sight. 
The inner disk should not be seen perfectly edge-on, as it would cause massive absorption in the stellar light, which is not observed. Since we know that the outer disk is just slightly inclined \citep[$i \approx 30^\circ$ according to][]{inclination}, we would expect a large tilt angle between the inner and outer disk. Such high misalignment was indeed observed  in the case of HD 142527 with a tilt angle of $70^\circ$ \citep{Marino2015}. In our case the size of the misaligned inner disk should agree with the range probed by our correlated spectrum measurements, that is, $\sim 1$~au in radius. 
The assumed geometry could be tested by high-resolution interferometric observations with different baseline position angles. Detailed modeling of this scenario is beyond the scope of this study.


\subsection{Silicate emission}
\label{sec:emission_radmodel}

We consider two main possibilities for the mid-infrared variability of the emission feature in the outer disk: Changing irradiation of a static disk by a variable central star, and changing disk structure around a non-variable star. 
In the first case, the irradiation of the outer disk may be due to either changing luminosity or a time-dependent shadowing close to the star. The stellar luminosity may change, for example, due to the time-variable accretion process. 
Due to a lack of simultaneous optical data along our mid-infrared observations, we cannot directly test the first case. Nevertheless, it would predict comparable flux changes in the inner and outer disks, as well as that the continuum and the silicate feature in the outer disk would change in the same manner. Our data do not show any of these correlations within the formal uncertainties, although they do not exclude the variable irradiation scenario entirely. 

\begin{figure}
\resizebox{\hsize}{!}{\includegraphics{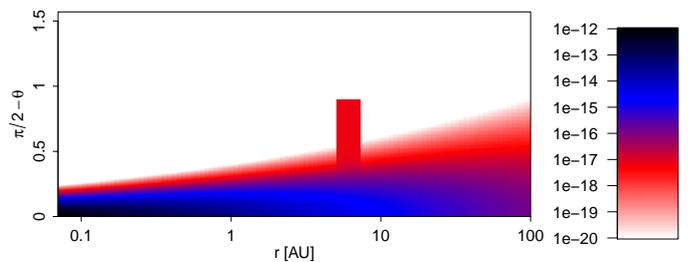}}
\caption{
Dust density distribution in spherical coordinates of our radiative transfer model including a cloud component between $5$~au and $7.5$~au. The density scale is $\mathrm{g\,cm}^{-3}$.
}
\label{fig:radmc_dens}
\end{figure}

As a next step, we explored changing disk structure as a source for the variability. To model the variability of the silicate emission in the outer disk outside $1-3$\,au, we include an additional small cloud above the disk, increasing the emitting area. The mass and opacity of the cloud determine the strength of the silicate feature.
The cloud is placed between $5$~au and $7.5$~au from the star (Fig.~{\ref{fig:radmc_dens}}), thus only influencing the outer disk spectrum. A smaller radius would lead to contamination of the inner disk spectrum. 

\begin{table}
	\small
\caption{Dust composition in the cloud component: $a_\mathrm{min}$ and $a_\mathrm{max}$ are the minimum and maximum diameters of the grains, respectively. $f_\mathrm{max}$ is the shape factor in the DHS-model \citep{Min2005dhs} and $\rho$ is the bulk density of the solid material. For each species we assume a power-law grain size distribution with an index of $-3.48$.
$f$ is the mass fraction in each species.
}
\begin{tabular}{llllll}
\toprule \toprule

Species                         
& $a_\mathrm{min}$ 
& $a_\mathrm{max}$
& $f_\mathrm{max}$
& $\rho$
& $f$
\\ 
& ($\mu$m) & ($\mu$m) & & (g\,cm$^{-3}$) & $\%$\\
\midrule
          amorphous carbon      &0.02  & 0.2  &0.0          &1.80   &15.1\\   amorphous olivine  &0.02  &185.6  &0.8          &3.71      &48.8
\\  amorphous pyroxene  &0.02  &185.6 &0.8          &3.20       &30.0
\\amorphous silica      &13  & 15  &0.8          &2.60          & 2.9
\\forsterite                &0.1  & 0.2  &1.0          &3.33    & 1.5
\\     
enstatite    &0.1 & 0.2 &1.0          &2.80     & 1.8
\\
\bottomrule
\end{tabular}
\label{tab:parampeterCloud}
\end{table}

In Sect.~\ref{sec:uncorr} we demonstrated that the silicate emission of DG Tau can be well described by the spectrum of comet Hale-Bopp (Fig. \ref{fig:uncorr_subtr}). Therefore we use a dust composition based on the modeling of this comet by \citet{Min2005}, but we fine-tuned some parameters to get a better match with the emission feature of DG Tau. To account for the effect of grain shape on optical properties, we used the distributed hollow sphere (DHS) model\footnote{The absorption and scattering coefficients as a functions of wavelength were computed using the software OpacityTool \citep{Min2005dhs,Toon1981}.} following \citet{Min2005dhs}, in which the shape effect is represented by the parameter $f_\mathrm{max}$.
We adjusted the dust composition, the grain size range, and the shape factor ($f_\mathrm{max}$) of amorphous carbon (see Table~\ref{tab:parampeterCloud}). 
For the dust species used in the model, the refractive indices as functions of wavelength are taken from
\citet[for amorphous carbon]{Jager1998carbon},
\citet[for amorphous olivine and amorphous pyroxene]{Dorschner1995},
\citet[for amorphous silica]{Henning1997},
\citet[for crystalline forsterite]{Zeidler2011},
\citet[for crystalline forsterite]{Fabian2001},
and
\citet[for crystalline enstatite]{Jager1998crystal}.

\begin{figure}
\resizebox{\hsize}{!}{\includegraphics{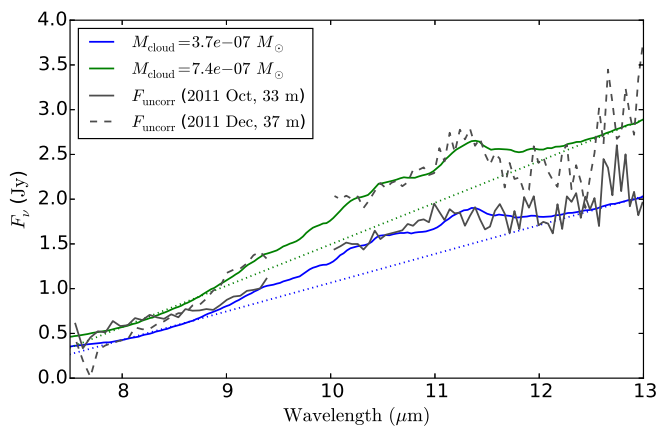}}
\caption{Mid-infrared outer disk ($>3$~au) spectra obtained from our radiative transfer model with different cloud masses (blue and green solid lines, for $3.7 \times 10^{-7} \mathrm{\,M}_\sun$ and $ 7.4 \times 10^{-7} \mathrm{\,M}_\sun$, respectively). Overplotted are MIDI observations of uncorrelated spectra from two epochs taken at similar baseline configurations (dashed and solid gray lines). Blue and green dotted lines show the continuum of the radiative transfer model.} \label{fig:radmc_em}
\end{figure}

We developed two variants of the model by using two different values for the cloud mass ($3.7 \times 10^{-7} \mathrm{\,M}_\sun$ and $7.4 \times 10^{-7} \mathrm{\,M}_\sun$). In both cases the cloud is optically thin for its own infrared emission but moderately optically thick for the optical stellar radiation ($\tau = 1$ and $2$, respectively). We adopted a gas-to-dust ratio of $100$. The basic parameters and the dust composition of the cloud are given in Tables~\ref{tab:radmc} and \ref{tab:parampeterCloud}, respectively.


The mid-infrared spectra of the outer disk ($>3$~au region) with the two different cloud masses are shown in Fig.~\ref{fig:radmc_em} (blue and green solid lines), along with two measured uncorrelated spectra (gray curves).  The shapes of the silicate features of the models and the observations are in general good agreement. We note that the model curves were not fitted to the data.  
We decomposed the resulting outer disk spectra in a similar fashion as was done for the observed uncorrelated spectra (Sect.~\ref{sec:uncorr}). We obtained $c_\mathrm{outer} = 0.4$ and $0.5$ for the strength of the silicate emission for the lower- and higher-mass models, respectively. This range is consistent with the data points in the lower half of Fig. \ref{fig:mjd_vs_c}. According to our simulations, increasing the cloud mass further cannot reproduce the stronger features, unless we change the geometry of the cloud by increasing its height. 
We also tested models with different cloud radii ($10$, $15$ and $20$~au). These models show that the dust temperature of the cloud is sensitive to the distance to the star. A cloud at $>10$~au would be too cold to produce considerable silicate emission, and therefore the cloud placed at $5$~au agrees best with most of the observations of the uncorrelated spectra.


Supporting the existence of dusty material high above the disk plane, \citet{Leinert1991} and \citet{Chen1992} detected extended emission around DG Tau with lunar occultation observations in the $K$ band. They argued that this emission can be caused by a $3.3 - 5$~au radius halo composed of optically thin dust which scatters stellar light in near-infrared. 
\citet{Chen1992} calculated that the total dust mass of the halo should be $3.5 \times 10^{-9} \mathrm{\,M}_{\sun}$, which may produce a line of sight absorption up to $A_\mathrm{V} = 2$. The calculated mass is in very good agreement with the dust mass of the cloud component ($M_\mathrm{cloud,\,dust} = 3.7-7.4 \times 10^{-9} \mathrm{\,M}_\sun$, assuming a gas-to-dust ratio of 100) in our radiative transfer model. Thus the results of \citet{Leinert1991} and \citet{Chen1992} seem to provide independent observational evidence for the existence of  dust high above the disk at $r \sim 3-5$~au scale, which can give rise to a variable silicate emission feature in our model.


Changing the vertical disk structure due to turbulence as a cause for mid-infrared variability was proposed theoretically by \citet{Turner2010}. They made magnetohydrodynamic calculations of a patch of disk orbiting a star. Over time intervals of a few orbits, the disk photosphere moves up and down due to magnetorotational turbulence. This process could provide a qualitative physical explanation for the origin of the cloud in our model, and could also explain that the mass of the cloud varies in time.

There is one issue, however, with this picture, that at distances where our uncorrelated spectra originate ($r>1-3$~au), the orbital periods and thus the typical dynamical timescales are in the range of a few years, while emission changes are observed on much shorter, monthly timescales (\cite{Bary_spitzer} and see Sect. \ref{sec:uncorr}). One possible explanation is that turbulence could also vertically lift up dust clouds at the inner edge of the dust disk. Then, according to \citet{Turner2010}, these nearby clouds can give rise to variability by casting time-varying shadows onto the more distant parts of the disk. This process can be effective on timescales consistent with the observations. However, the emission of these variable inner clouds would contribute to the correlated spectra. Thus this scenario would predict an anti-correlation between the correlated and uncorrelated flux levels, which is not observed in our data at the level of the formal uncertainties. To be able to check the possibility of disk shadowing we would need larger multi-epoch data sets at similar baseline configurations.




One remaining point concerning our observations is the observed unusual radial distribution of the crystalline material. Unless the crystalline material has been produced locally in shock fronts \citep{Harker2002}, an interesting question is how it was transported to the outer regions of the mid-infrared emitting disk, where the temperature is too low to produce crystals locally \citep{Colangeli2003}. Outflows, surface flows, and disk winds may be capable of transferring the crystalline dust from hot inner parts to the outer disk. \citet{Lopez2014} observed the semi-forbidden lines of CII], SiII], and FeII] in the $231-234$~nm wavelength range for a sample of T Tauri stars, including DG Tau. They observed that DG Tau has blueshifted line centroids, which are indeed characteristic of outflows. Additionally, the presence of the jet is also a sign for the existence of outward material transport. The fact that we do not detect crystalline dust in the inner disk may suggest that the crystal production is not continuous, but linked to episodes of higher activity, after which the fresh crystals are transported outwards, as in the case of EX Lupi \citep{Abraham2009,Juhasz2012}.


\section{Summary}
\label{sec:concl}

In this study we analyzed multi-epoch mid-infrared interferometric data set of DG Tau, a classical T Tauri-type star in the Taurus star forming region. At mid-infrared wavelengths, most of the light comes from the inner few au part of the circumstellar disk. We obtained VLTI/MIDI observations between 2011 and 2014, enabling us to investigate the temporal variability of the source. In the interferometric observations, we obtained correlated and uncorrelated spectra, dominated by the inner and outer region of the disk (inside and outside a radius of $1-3$~au, depending on the resolution), respectively. Our main result is that we found a striking difference between the inner and outer disk spectral features (a switch between absorption and emission) which is highly unusual among T Tauri stars. The detailed conclusions of this study are as follows:
\begin{itemize}
\item Based on the correlated spectra, the half-light radius of the innermost part of the system ($r< 3$~au), measured at $10.7~\mu$m, is $r_\mathrm{e} = 0.73 \pm 0.10~ \mathrm{au}$.

\item 
The sizes for the whole mid-infrared-emitting region (including the contribution from the outer, resolved part of the disk) determined  for each epoch show a decreasing trend with time. Between 2011 and 2014 the half-light radius (at $10.7~\mu$m) decreased from $r^V_\mathrm{e} = 1.15 \pm 0.28$~au to $0.73 \pm 0.12$~au. The decrease is mainly due to brightness changes in the outer disk.

\item The correlated spectra exhibit $10~\mu$m absorption features related to small amorphous silicate grains. The uncorrelated spectra show silicate features in emission. The shape of the emission features indicates the presence of crystalline grains. 

\item We could fit the spectra with a combination of a linear continuum and a normalized spectral template of the silicate feature. We used a spectrum of the ISM towards the galactic center to model the silicate absorption in the inner disk spectra. For the outer disk spectra we used the spectrum of comet Hale-Bopp as a template.

\item The average amplitude of the emission in the outer disk spectra
is $0.63$~Jy, twice as large as the inner absorption. The silicate feature in the outer disk exhibits temporal variability, while in the inner disk the amplitude of the absorption has a spatial dependence, which means that by zooming into the disk, the amplitude of the absorption increases.

\item We performed a radiative transfer modeling and could describe the observed SED of the system, assuming a flared continuous disk between $0.01$ au and $100$ au.

\item We discussed four scenarios for the origin of the absorption in the inner disk: a cold obscuring envelope, an accretion-heated inner disk, a temperature inversion on the disk surface and a misaligned inner geometry. The first scenario can partly explain the silicate absorption, although it is required that in the inner disk the intrinsic silicate emission should be much weaker than in the outer disk. Accretion heating from the mid-plane in the inner disk can cause a weakened intrinsic emission. The third scenario might be significant near the sublimation radius. Thus, the observed characteristics of the absorption profile may be explained by the combined effect of the first three processes.
Our last scenario can be an alternative, although a very special geometry is needed.  


\item The variable crystalline emission in the outer disk could be modeled by a cloud high above the disk plane at $\sim 5$~au from the star. Changing the mass of this cloud could account for the observed variability of the emission feature, although timescales shorter than predicted by this model were also observed. 

\end{itemize}

A deeper understanding of this unique system would require  simultaneous observations in the optical and infrared wavelengths, including high-resolution infrared interferometric data (GRAVITY, MATISSE). This kind of observation could reveal the effect of changes in the disk irradiation by the central star and the accretion dynamics on the silicate feature properties. 

\begin{acknowledgements}
            This work was supported by the Hungarian Research Fund OTKA grant K101393 and by the Momentum grant of the MTA CSFK Lend\"ulet Disk Research Group (LP2014-6). The authors are thankful for the support of the Fizeau Exchange Visitor Program (OPTICON/FP7) through the European Interferometry Initiative (EII) funded by WP14 OPTICON/FP7 (2013--2016, grant number 312430).
We would like to thank Michiel Min for useful discussions. Also, we thank our anonymous referee for helpful comments.

\end{acknowledgements}


\bibliographystyle{aa}
\bibliography{ref_DGTau,dust}





\appendix
\section{Uncertainties of the fitted parameters}
\label{app:err}

In Sect.~\ref{sec:correlated} and \ref{sec:uncorr} we presented our fits to the correlated and uncorrelated MIDI spectra. In this appendix, we describe our method to determine the uncertainties of the fitted parameters.
First, we created $1000$ synthetic spectra by data resampling in every epoch and ran the same fitting procedure on them. Then the uncertainties of the parameters were taken as the standard deviation of the best parameter values. To create the synthetic spectra, we attempted to disentangle the random and systematic effects in the errors of the measured spectra. We smoothed the datasets with a rolling window of width $7$ points. This width was chosen in a way to represent the real resolution of the MIDI low resolution spectrum ($\lambda/\Delta \lambda \sim 30$). Then the random errors were calculated as the standard deviation of the measured points around the smoothed curve within this window. The systematic errors were calculated assuming that the measured uncertainties are the squared sum of the random and systematic errors. 

When creating the synthetic spectra, at every wavelength value 
we randomly chose a point from a normally distributed pool of data with a standard deviation given by the random error and a mean given by the measurement at that wavelength. To account for the systematic error the whole spectrum was shifted in flux density. The amount and direction of this offset was determined from the systematic errors in the following way: a random value was chosen at the wavelength $10.7$\,$\mu$m from a normal distribution around zero with a standard deviation given by the systematic error at this wavelength. The offset values at the other wavelengths were calculated using the proportionality shown by the systematic errors (systematic errors increase with wavelength). 
When fitting synthetic spectra, we found that the synthetic best-fit parameters have essentially a random normal distribution around the real best-fit values.

\listofobjects
\end{document}